\documentclass[letterpaper]{IEEEtran}
\usepackage{cite}
\usepackage{epsfig}
\usepackage{amssymb}
\usepackage{mathrsfs}
\usepackage{times}
\begin{document}
%*******************************************************************************
%*******************************************************************************
\title{Coherent Detection of Turbo-Coded OFDM Signals Transmitted
through Frequency Selective Rayleigh Fading Channels with Receiver Diversity
and Increased Throughput}
\author{K. Vasudevan
\thanks{
This work is supported by the India-UK Advanced Technology Center (IU-ATC)
of Excellence in Next Generation Networks, Systems and Services under grant
SR/RCUK-DST/IUATC Phase 2/2012-IIT K (C), sponsored by DST-EPSRC.}

\thanks{This work was presented as an Invited Talk
at the International Federation of Nonlinear Analysts (IFNA)
World Congress, Greece, June 2012 and
partly published in ISPCC, Sept. 2013, Shimla and ITST Nov. 2013, Finland.}

\thanks{The author is with the Dept. of EE, IIT Kanpur.
 Email: vasu@iitk.ac.in}
}
%*******************************************************************************
\maketitle
%*******************************************************************************
\thispagestyle{empty}
%*******************************************************************************
\begin{abstract}
In this work, we discuss techniques for coherently detecting turbo coded
orthogonal frequency division multiplexed (OFDM) signals, transmitted through
frequency selective Rayleigh (the magnitude of each channel tap is Rayleigh
distributed) fading channels having a uniform power delay
profile. The channel output is further distorted
by a carrier frequency and phase offset, besides additive white Gaussian noise
(AWGN). A new frame structure for OFDM, consisting of a known preamble, cyclic
prefix, data and known postamble is proposed, which has a higher throughput
compared to the earlier work. A robust turbo decoder is proposed, which
functions effectively over a wide range of signal-to-noise ratio (SNR).
Simulation results show that it is possible to achieve a bit-error-rate (BER)
of $10^{-5}$ at an SNR per bit as low as 8 dB and throughput of 82.84\%, using
a single transmit and two receive antennas. We also demonstrate that the
practical coherent receiver requires just about 1 dB more power compared to
that of an ideal coherent receiver, to attain a BER of $10^{-5}$. The key
contribution to the good performance of the practical coherent receiver is
due to the use of a long preamble (512 QPSK symbols), which is
perhaps not specified in any of the current wireless communication standards.
We have also shown from computer simulations that, it is possible to
obtain even better BER performance, using a better code.
A simple and approximate Cram{\' e}r-Rao bound on the
variance of the frequency offset estimation error for coherent detection, is
derived.
The proposed algorithms are well suited for implementation on a DSP-platform.
\end{abstract}
%*******************************************************************************
%*******************************************************************************
\IEEEpeerreviewmaketitle
%*******************************************************************************
\section{Introduction}
\label{Sec:Intro}
Future wireless communication standards aim to push the existing data-rates
higher. This can only be achieved with the help of coherent communications,
since they give the lowest bit-error-rate (BER) performance for a given
signal-to-noise ratio (SNR). Conversely, they require the lowest SNR to
attain a given BER, resulting in enhanced battery life. If we look at a
mobile, it indicates a typical received signal strength equal to $-100$ dBm
($10^{-10}$ mW). However this
is not the signal-to-noise ratio! Therefore, the question
is: What is the operating SNR of the mobiles? Would it be possible to achieve
the same performance by transmitting at a lower power? The recent advances in
cooperative communications has resulted in low complexity solutions, that are
not necessarily power efficient \cite{Hanzo_2011,Zhang_2013}. In fact, it is
worth quoting the following from \cite{Hanzo_2012}:
%*******************************************************************************
\begin{enumerate}
 \item {\it The Myth: Sixty years of research following Shannon's
       pioneering paper has led to telecommunications
       solutions operating arbitrarily close to the channel
       capacity--``flawless telepresence'' with zero error is
       available to anyone, anywhere, anytime across the globe.}
 \item {\it The Reality: Once we leave home or the office, even
       top of the range iPhones and tablet computers fail to
       maintain ``flawless telepresence'' quality. They also fail
       to approach the theoretical performance predictions.
       The 1000-fold throughput increase of the best third-
       generation (3G) phones over second-generation (2G)
       GSM phones and the 1000-fold increased teletraffic
       predictions of the next decade require substantial
       further bandwidth expansion toward ever increasing
       carrier frequencies, expanding beyond the radio-
       frequency (RF) band to optical frequencies, where
       substantial bandwidths are available.}
\end{enumerate}
%*******************************************************************************
The transmitter and receiver algorithms proposed in this paper and
in \cite{Vasu13,Umesh13} are well suited for implementation on a DSP processor
or hardwired and may perhaps not require quantum computers, as mentioned in
\cite{Hanzo_2012}. The reader is also referred to the brief commentary
on channel estimation and synchronization in page 1351 and also to the
noncoherent schemes in page 1353 of \cite{Hanzo_2011}, which clearly state
that cooperative communications avoid coherent receivers due to
complexity.

Broadly speaking, the wireless communication device needs to have the
following features:
%*******************************************************************************
\begin{enumerate}
 \item maximize the bit-rate
 \item minimize the bit-error-rate
 \item minimize transmit power
 \item minimize transmission bandwidth
\end{enumerate}
%*******************************************************************************
A rather disturbing trend in the present day wireless communication systems is
to make the physical layer very simple and implement it in hardware, and
allot most of the computing resources to the application layer, e.g., for
internet surfing, video conferencing etc. While hardware implementation of
the physical layer is not an issue, in fact, it may even be preferred over
software implementation in some situations, the real cause for concern is the
tendency to make it ``simple'', at the cost of BER performance. Therefore,
the questions are:
%*******************************************************************************
\begin{enumerate}
 \item was signal processing for coherent communications given a
       chance to prove itself, or was it ignored straightaway, due to
       ``complexity'' reasons?
 \item are the present day single antenna wireless transceivers,
       let alone multi-antenna systems, performing anywhere near channel
       capacity?
\end{enumerate}
%*******************************************************************************
This paper demonstrates that coherent receivers need not be
restricted to textbooks alone, in fact they can be implemented with
linear (not exponential) complexity. The need of the hour is a paradigm
shift in the way the wireless communication systems are implemented.

In this article, we dwell on coherent receivers based on orthogonal frequency
division multiplexing (OFDM),
since it has the ability to mitigate intersymbol interference (ISI) introduced
by the frequency selective fading channel
\cite{Bingham90,Vasu_Book10,Hanzo_OFDM_Primer}. The ``complexity'' of
coherent detection can be overcome by means of parallel processing, for
which there is a large scope. We wish to emphasize
that this article presents a proof-of-concept, and is hence not constrained
by the existing standards in wireless communication.
We begin by first outlining the tasks of a coherent receiver. Next, we scan
the literature on each of these tasks to find out the state-of-the-art, and
finally end this section with our contributions.

The basic tasks of the coherent receiver would be:
%*******************************************************************************
\begin{enumerate}
 \item To correctly identify the start of the (OFDM) frame (SoF), such that
       the probability of false alarm (detecting an OFDM
       frame when it is not present) or equivalently the probability of
       erasure/miss (not detecting the OFDM frame when it is
       present) is minimized. We refer to this step as timing synchronization.
 \item To estimate and compensate the carrier frequency offset (CFO),
       since OFDM is known to be sensitive to CFO. This task is referred to as
       carrier synchronization.
 \item To estimate the channel impulse/frequency response.
 \item To perform (coherent) turbo decoding and recover the data.
\end{enumerate}
%*******************************************************************************
To summarize, a coherent receiver at the physical layer ensures that the
medium access control (MAC) is not burdened by frequent requests
for retransmissions.

A robust timing and frequency synchronization for OFDM signals transmitted
through frequency selective AWGN channels is presented in \cite{Cox97}.
Timing synchronization in OFDM is addressed in \cite{Beek95_2,Landstrom02,
Cheon03,Ren05,Kang08}.
Various methods of carrier frequency synchronization for OFDM are
given in \cite{Baum98,Garcia01,Bradaric03,Kuo05,Lin_Chen_05,Ahn07,Henkel07}.
Joint timing and CFO estimation is discussed in
\cite{Tufvesson99,Minn03,Zhang05,Ziabari2011,Tanda2013,Salcedo2013}.

Decision directed coherent detection of
OFDM in the presence of Rayleigh fading is treated in \cite{Frenger99}.
A factor graph  approach to the iterative (coherent) detection of OFDM
in the presence of carrier frequency offset and phase noise
is presented in \cite{Merli08}. OFDM detection in the presence of
intercarrier interference (ICI) using block whitening is discussed in
\cite{Wang12}. In \cite{Chen2013}, a turbo receiver is proposed for detecting
OFDM signals in the presence of ICI and inter antenna interference.

Most flavors of
the channel estimation techniques discussed in the literature are done in the
frequency domain, using pilot symbols at regular intervals in the
time/frequency grid
\cite{Beek95,Edfors98,Puri02,Ribeiro08,Kinjo08}. Iterative joint channel
estimation and multi-user detection for multi-antenna OFDM is discussed in
\cite{Jiang07}. Noncoherent detection of coded OFDM in the {\it absence of
frequency offset\/} and assuming that the channel frequency response to be
constant over a block of symbols, is considered in \cite{Fischer01}.
Expectation maximization (EM)-based joint channel estimation and exploitation
of the diversity gain from IQ imbalances is addressed in \cite{Marey12}.

Detection of OFDM signals, in the context of spectrum sensing for cognitive
radio,
is considered in \cite{Kamalian12,Turunen12}. However, in both these papers,
the probability of false alarm is quite high (5\%).

In \cite{Peng06}, discrete cosine transform (DCT)
based OFDM is studied in the presence of frequency offset and noise, and
its performance is compared with the discrete Fourier transform (DFT) based
OFDM. It is further shown in \cite{Peng06} that the performance of DFT-OFDM
is as good as DCT-OFDM, for small frequency offsets.

A low-power OFDM
implementation for wireless local area networks (WLAN) is addressed in
\cite{Yu12}. OFDM is a suggested modulation technique for digital video
broadcasting \cite{Jeanclaude95,Reimers98}. It has also been proposed for
optical communications \cite{Takahashi09}.

The novelty of this work lies in the use of a filter that is matched to the
preamble, to acquire timing
synchronization \cite{Vasu08,Vasu_SIVP10} (start-of-frame (SoF) detection).
Maximum likelihood (ML) channel estimation using the preamble is performed.
This approach does not require any knowledge of the channel and noise
statistics.

The main contributions of this paper are the following:
%*******************************************************************************
\begin{enumerate}
 \item It is shown that, for a sufficiently long preamble, the variance of the
       channel
       estimator proposed in eq. (28) of \cite{Vasu13} approaches zero.
 \item A known postamble is used to accurately estimate the residual frequency
       offset for large data lengths, thereby increasing the throughput
       compared to \cite{Vasu13,Umesh13}.
 \item Turbo codes are used to attain BER performance closer to
       channel capacity compared to any other earlier work in the open
       literature, for channels having a uniform power delay profile (to the
       best of the authors knowledge, there is no similar
       work on the topic of this paper, other than \cite{Vasu13,Umesh13}).
 \item A robust turbo decoder is proposed, which performs effectively over
       a wide range of SNR (0 -- 30 dB).
 \item While most papers in the literature try to attain the channel capacity
       for a given SNR, this work tries to attain the minimum SNR for
       error-free transmission, for a given channel capacity.
\end{enumerate}
%*******************************************************************************
In a multiuser scenario, the suggested technique is OFDM-TDMA. The uplink
and downlink may be implemented using time division duplex (TDD) or
frequency division duplex (FDD) modes.

This paper is organized as follows. Section~\ref{Sec:Sys_Model} describes the
system model. The receiver algorithms are presented in
section~\ref{Sec:Receiver}. The bit-error-rate (BER) results from
computer simulations are given in section~\ref{Sec:Sim_Results}. Finally, in
section~\ref{Sec:Conclude}, we discuss the conclusions and future work.

%*******************************************************************************
\section{System Model}
\label{Sec:Sys_Model}
%*******************************************************************************
\begin{figure}[tbh]
\centering
\input{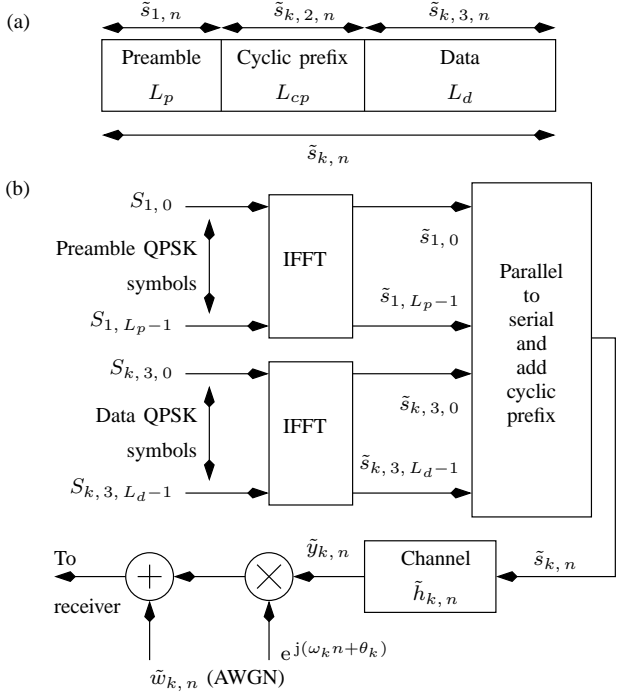}
\caption{(a) The frame structure. (b) System model. $k$ denotes the frame
         index and $n$ denotes the time index in a given frame.}
\label{Fig:Frame}
\end{figure}
%*******************************************************************************
We assume that the data to be transmitted is organized into frames, as
depicted in Figure~\ref{Fig:Frame}.
The frame consists of a known preamble of length $L_p$ symbols, a cyclic
prefix of length $L_{cp}$, followed by data of length $L_d$ symbols.
Thus, the total length of the frame is
%*******************************************************************************
\begin{eqnarray}
\label{Eq:Pap8_Eq1}
L = L_p + L_{cp} + L_d.
\end{eqnarray}
%*******************************************************************************
Let us assume a channel span equal to $L_h$. The channel span assumed by the
receiver is $L_{hr}$ ($> L_h$). The length of the cyclic prefix is
\cite{Vasu_Book10}:
%*******************************************************************************
\begin{eqnarray}
\label{Eq:Pap8_Eq2}
L_{cp} = L_{hr} - 1.
\end{eqnarray}
%*******************************************************************************
Throughout the manuscript, we use tilde  to denote complex quantities.
However, complex (QPSK) symbols will be denoted without a tilde e.g.
$S_{1,\, n}$. Boldface letters denote vectors or matrices.
The channel coefficients $\tilde h_{k,\, n}$ for the $k^{th}$ frame are
$\mathscr{CN}(0,\, 2\sigma^2_f)$ and independent over time ($n$), that is:
%*******************************************************************************
\begin{eqnarray}
\label{Eq:Pap8_Eq3}
\frac{1}{2}
 E
\left[
\tilde h_{k,\, n}
\tilde h_{k,\, n-m}^*
\right] = \sigma^2_f \delta_K(m)
\end{eqnarray}
%*******************************************************************************
where ``*'' denotes complex conjugate and $\delta_K(\cdot)$ is the
Kronecker delta function. This implies a uniform channel power delay profile.
The channel is assumed to be quasi-static, that is $\tilde h_{k,\, n}$
is time-invariant over one frame and varies independently from frame-to-frame,
that is
%*******************************************************************************
\begin{eqnarray}
\label{Eq:Pap8_Eq4}
\frac{1}{2}
 E
\left[
\tilde h_{k,\, n}
\tilde h_{j,\, n}^*
\right] = \sigma^2_f \delta_K(k-j)
\end{eqnarray}
%*******************************************************************************
where $k$ and $j$ denote the frame indexes. The AWGN noise samples
$\tilde w_{k,\, n}$ for the $k^{th}$ frame at time $n$ are
$\mathscr{CN}(0,\, 2\sigma^2_w)$. The frequency offset
$\omega_k$ for the $k^{th}$ frame is uniformly distributed over
$[-0.04,\, 0.04]$ radian \cite{Minn03}.
The phase offset $\theta_k$ for the $k^{th}$ frame is uniformly distributed
over $[0,\, 2\pi)$.
Both $\omega_k$ and $\theta_k$ are fixed for a frame and vary randomly
from frame-to-frame.

Note that:
%*******************************************************************************
\begin{eqnarray}
\label{Eq:Pap8_Eq4_1}
\tilde s_{1,\, n} & = & \frac{1}{L_p}
                        \sum_{i=0}^{L_p-1}
                         S_{1,\, i}
                        \mathrm{e}^{\,\mathrm{j}\, 2\pi ni/L_p}
                        \quad \mbox{for $0\le n \le L_p-1$}  \nonumber  \\
\tilde s_{k,\, 3,\, n}
                  & = & \frac{1}{L_d}
                        \sum_{i=0}^{L_d-1}
                         S_{k,\, 3,\, i}
                        \mathrm{e}^{\,\mathrm{j}\, 2\pi ni/L_d}
                        \quad \mbox{for $0\le n \le L_d-1$}  \nonumber  \\
\tilde s_{k,\, 2,\, n}
                  & = & \tilde s_{k,\, 3,\, L_d-L_{cp}+n}
                        \quad \mbox{for $0 \le n \le L_{cp}-1$}.
\end{eqnarray}
%*******************************************************************************
We assume $S_{k,\, 3,\, i}\in \pm 1 \pm \mathrm{j}$. Since we require:
%*******************************************************************************
\begin{eqnarray}
\label{Eq:Pap8_Eq4_2}
 E
\left[
\left|
\tilde s_{1,\, n}
\right|^2
\right] & = &
 E
\left[
\left|
\tilde s_{k,\, 3,\, n}
\right|^2
\right] = 2/L_d \stackrel{\Delta}{=} \sigma^2_s
\end{eqnarray}
%*******************************************************************************
we must have $S_{1,\, i}\in \sqrt{L_p/L_d}\,(\pm 1 \pm \mathrm{j})$. In other
words, the average power of the preamble part must be equal to the average
power of the data part.

The received signal for the $k^{th}$ frame can be written
as (for $0 \le n \le L+L_h-2$):
%*******************************************************************************
\begin{eqnarray}
\label{Eq:Pap8_Eq5}
\tilde r_{k,\, n} & = & \left(
                        \tilde s_{k,\, n} \star \tilde h_{k,\, n}
                        \right)\,
                        \mathrm{e}^{\,\mathrm{j}(\omega_k n+\theta_k)} +
                        \tilde w_{k,\, n}                 \nonumber  \\
                  & = & \tilde y_{k,\, n}
                        \mathrm{e}^{\,\mathrm{j}(\omega_k n+\theta_k)} +
                        \tilde w_{k,\, n}
\end{eqnarray}
%*******************************************************************************
where ``$\star$'' denotes convolution and
%*******************************************************************************
\begin{eqnarray}
\label{Eq:Pap8_Eq5_1}
\tilde y_{k,\, n} = \tilde s_{k,\, n} \star \tilde h_{k,\, n}.
\end{eqnarray}
%*******************************************************************************
The set of received samples can be denoted by the vector:
%*******************************************************************************
\begin{eqnarray}
\label{Eq:Pap8_Eq5_2}
\tilde\mathbf{r}_k =
\left[
\begin{array}{ccc}
\tilde r_{k,\, 0} & \ldots & \tilde r_{k,\, L+L_h-2}
\end{array}
\right].
\end{eqnarray}
%*******************************************************************************

%*******************************************************************************
\section{Receiver}
\label{Sec:Receiver}
%*******************************************************************************
In this section we discuss the key receiver algorithms, namely, start of frame
(SoF), coarse/fine frequency offset, channel and noise variance estimation and
finally data detection.
%*******************************************************************************
\subsection{Start of Frame and Coarse Frequency Offset Estimation}
\label{SSec:Frame_Detection}
%*******************************************************************************
Let us assume that for the $k^{th}$ frame, the channel impulse response
is known at the receiver. The channel length assumed by the receiver is
$L_{hr}(> L_h)$ such that the first $L_h$ coefficients are identical to the
channel coefficients and the remaining $L_{hr}-L_h$ coefficients are zeros.
Define the $m^{th}$
($0 \le m \le L_{cp}+L_d+L_h+L_{hr}-2$) received vector as:
%*******************************************************************************
\begin{eqnarray}
\label{Eq:Pap8_Eq5_3}
\tilde\mathbf{r}_{k,\, m} =
\left[
\begin{array}{ccc}
\tilde r_{k,\, m} & \ldots & \tilde r_{k,\, m+L_p-L_{hr}}
\end{array}
\right].
\end{eqnarray}
%*******************************************************************************
The steady-state\footnote{By steady-state we mean that all the channel
coefficients are involved in the convolution to generate $\tilde y_{k,\, n}$
in (\ref{Eq:Pap8_Eq5_1})} preamble part of the transmitted signal appearing at
the channel output can be represented by a vector:
%*******************************************************************************
\begin{eqnarray}
\label{Eq:Pap8_Eq5_4}
\tilde\mathbf{y}_{k,\, 1} =
\left[
\begin{array}{ccc}
\tilde y_{k,\, L_{hr}-1} & \ldots & \tilde y_{k,\, L_p-1}
\end{array}
\right].
\end{eqnarray}
%*******************************************************************************
The non-coherent maximum likelihood (ML) rule for frame detection
can be stated as \cite{Vasu_Book10}: Choose that time as the start of frame
and that frequency $\hat\omega_k$, which jointly maximize the conditional pdf:
%*******************************************************************************
\begin{eqnarray}
\label{Eq:Pap8_Eq6}
\max_{m,\,\hat\omega_k}
\int_{\theta_k=0}^{2\pi}
 p
\left(
\tilde\mathbf{r}_{k,\, m}|\tilde\mathbf{y}_{k,\, 1},\hat\omega_k,\theta_k
\right) p(\theta_k)\, d\theta_k.
\end{eqnarray}
%*******************************************************************************
substituting for the joint pdf and $p(\theta_k)$ and defining
%*******************************************************************************
\begin{eqnarray}
\label{Eq:Pap8_Eq6_1}
L_1=L_p-L_{hr}+1
\end{eqnarray}
%*******************************************************************************
we get:
%*******************************************************************************
\begin{eqnarray}
\label{Eq:Pap8_Eq7}
\lefteqn{
\max_{m,\,\hat\omega_k}
\frac{1}{2\pi}
\frac{1}{(2\pi\sigma^2_w)^{L_1}}}                        \nonumber  \\
&   &
\int_{\theta=0}^{2\pi}
\exp
\left(
\frac{
-
\sum_{i=0}^{L_1-1}
\left|
\tilde r_{m+i}-\tilde y_{k,\, L_{hr}-1+i}
\,
\mathrm{e}^{\,\mathrm{j}(\hat\omega_k i+\theta)}
\right|^2
}{2\sigma^2_w}
\right)                                                  \nonumber  \\
&   & \mbox{ } \times \, d\theta.
\end{eqnarray}
%*******************************************************************************
where
%*******************************************************************************
\begin{eqnarray}
\label{Eq:Pap8_Eq8}
\theta = \hat\omega_k(L_{hr}-1) + \theta_k
\end{eqnarray}
%*******************************************************************************
incorporates the phase accumulated by the frequency offset over the first
$L_{hr}-1$ samples, besides the initial phase $\theta_k$. Observe that $\theta$
is also uniformly distributed in $[0,\, 2\pi)$.

\begin{figure}[tbh]
\centering
\input{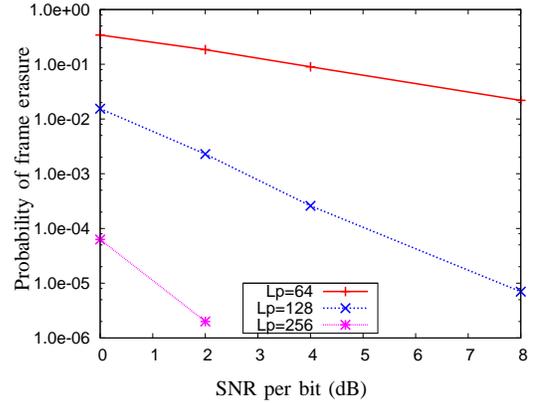}
\caption{Probability of frame erasure as a function of the preamble length
         $L_p$. \copyright\, 2013 IEEE. Reprinted, with permission, from
         \cite{Vasu13}}
\label{Fig:DMT4_Erase}
\end{figure}

One of the terms in the exponent is:
%*******************************************************************************
\begin{eqnarray}
\label{Eq:Pap8_Eq9}
\frac{\sum_{i=0}^{L_1-1}|\tilde r_{m+i}|^2}{2\sigma^2_w}
\end{eqnarray}
%*******************************************************************************
is approximately proportional to the average received signal power, for large
values of $L_p$ and $L_p \gg L_{hr}$, and is hence (approximately) independent
of $m$ and $\theta$. The other exponential term
%*******************************************************************************
\begin{eqnarray}
\label{Eq:Pap8_Eq10}
\frac{\sum_{i=0}^{L_1-1}|\tilde y_{k,\, L_{hr}-1+i}|^2}{2\sigma^2_w}
\end{eqnarray}
%*******************************************************************************
is clearly independent of $m$ and $\theta$. Therefore we are only left with
(ignoring constants):
%*******************************************************************************
\begin{eqnarray}
\label{Eq:Pap8_Eq11}
\hspace*{0.1in}
\lefteqn{
\max_{m,\,\hat\omega_k}
\frac{1}{2\pi}}                                          \nonumber  \\
&   &
\hspace*{-0.15in}
\int_{\theta=0}^{2\pi}
\exp
\left(
\frac{
\Re
\left
\{
\sum_{i=0}^{L_1-1}
 2
\tilde r_{m+i}\,\tilde y_{k,\, L_{hr}-1+i}^*
\,
\mathrm{e}^{-\mathrm{j}(\hat\omega_k i+\theta)}
\right
\}
}{2\sigma^2_w}
\right)                                                  \nonumber  \\
&   & \mbox{ } \times \, d\theta
\end{eqnarray}
%*******************************************************************************
which simplifies to \cite{Vasu_Book10}:
%*******************************************************************************
\begin{eqnarray}
\label{Eq:Pap8_Eq12}
\max_{m,\,\hat\omega_k}
I_0
\left(
\frac{A_{m,\,\hat\omega_k}}
{2\sigma^2_w}
\right)
\end{eqnarray}
%*******************************************************************************
where $I_0(\cdot)$ is the modified Bessel function of the zeroth-order and
%*******************************************************************************
\begin{eqnarray}
\label{Eq:Pap8_Eq13}
A_{m,\,\hat\omega_k} = \left|
                       \sum_{i=0}^{L_1-1}
                        2
                       \tilde r_{m+i}\,\tilde y_{k,\, L_{hr}-1+i}^*
                       \,
                       \mathrm{e}^{-\mathrm{j}\,\hat\omega_k i}
                       \right|.
\end{eqnarray}
%*******************************************************************************
Noting that $I_0(x)$ is a monotonic function of $x$ and ignoring constants,
the maximization in (\ref{Eq:Pap8_Eq12}) simplifies to:
%*******************************************************************************
\begin{eqnarray}
\label{Eq:Pap8_Eq14}
\max_{m,\,\hat\omega_k}
                       \left|
                       \sum_{i=0}^{L_1-1}
                       \tilde r_{m+i}\,\tilde y_{k,\, L_{hr}-1+i}^*
                       \,
                       \mathrm{e}^{-\mathrm{j}\,\hat\omega_k i}
                       \right|.
\end{eqnarray}
%*******************************************************************************
Observe that (\ref{Eq:Pap8_Eq14}) resembles the operation of demodulation and
matched filtering. The ideal outcome of (\ref{Eq:Pap8_Eq14}) to estimate
the SoF and frequency offset is:
%*******************************************************************************
\begin{eqnarray}
\label{Eq:Pap8_Eq14_1}
m            & = & L_{hr} -1    \nonumber  \\
\hat\omega_k & = & \omega_k.
\end{eqnarray}
%*******************************************************************************
In practice, the receiver has only the estimate of the
channel ($\hat h_{k,\, n}$), hence $\tilde y_{k,\, n}$ must be replaced by
$\hat y_{k,\, n}$, where
%*******************************************************************************
\begin{eqnarray}
\label{Eq:Pap8_Eq15}
\hat y_{k,\, n} = \tilde s_{1,\, n} \star \hat h_{k,\, n}
\end{eqnarray}
%*******************************************************************************
is the preamble convolved with the channel estimate.
When $\hat h_{k,\, n}$ is not available, we propose a heuristic method
of frame detection as follows:
%*******************************************************************************
\begin{eqnarray}
\label{Eq:Pap8_Eq16}
\max_{m,\,\hat\omega_k}
                       \left|
                       \sum_{i=0}^{L_p-1}
                       \tilde r_{m+i}\,\tilde s_{1,\, i}^*
                       \,
                       \mathrm{e}^{-\mathrm{j}\,\hat\omega_k i}
                       \right|
\end{eqnarray}
%*******************************************************************************
where again $\tilde s_{1,\, i}$ denotes the preamble as shown in
Figure~\ref{Fig:Frame}. The ideal outcome of (\ref{Eq:Pap8_Eq16}) is:
%*******************************************************************************
\begin{eqnarray}
\label{Eq:Pap8_Eq17}
0            & \le
                 & m \le L_h - 1    \nonumber  \\
\hat\omega_k & = & \omega_k
\end{eqnarray}
%*******************************************************************************
depending on which channel coefficient has the maximum magnitude. In practical
situations, one also needs to look at the ratio of the peak-to-average
power of (\ref{Eq:Pap8_Eq16}) to estimate the SoF \cite{Vasu_SIVP10}. When $m$
lies outside the range in (\ref{Eq:Pap8_Eq17}), the frame is declared as
erased (lost). The probability of frame erasure as a function of the preamble
length is shown in Figure~\ref{Fig:DMT4_Erase}. Observe that for $L_p=512$, the
probability of erasure is less than $10^{-6}$ and is hence not plotted.
%*******************************************************************************
%*******************************************************************************

The coarse frequency offset estimate $\hat\omega_k$ is obtained by dividing the
interval $[-0.04,\, 0.04]$ radian into $B_1$ frequency bins and selecting
that bin which maximizes (\ref{Eq:Pap8_Eq16}).

%*******************************************************************************
\subsection{Channel Estimation}
\label{SSec:Channel_Est}
%*******************************************************************************
Here, we focus on maximum likelihood (ML) channel estimation. We assume that
the SoF has been estimated using (\ref{Eq:Pap8_Eq16}) with outcome $m_0$
($0 \le m_0 \le L_h-1$) and the frequency offset has been perfectly canceled.
Define
%*******************************************************************************
\begin{eqnarray}
\label{Eq:Pap8_Eq18_0}
m_1 = m_0 + L_h -1.
\end{eqnarray}
%*******************************************************************************

The steady-state,
preamble part of the received signal for the $k^{th}$ frame can be written as:
%*******************************************************************************
\begin{eqnarray}
\label{Eq:Pap8_Eq18}
\tilde\mathbf{r}_{k,\, m_1} = \tilde\mathbf{s}_1 \tilde\mathbf{h}_k +
                              \tilde\mathbf{w}_{k,\, m_1}
\end{eqnarray}
%*******************************************************************************
where
%*******************************************************************************
\begin{eqnarray}
\label{Eq:Pap8_Eq19}
\tilde\mathbf{r}_{k,\, m_1}
& = &
\left[
\begin{array}{ccc}
\tilde r_{k,\, m_1} & \ldots  & \tilde r_{k,\, m_1+L_p-L_{hr}}
\end{array}
\right]^T                             \nonumber  \\
&   & \mbox{ }
      [(L_p-L_{hr}+1)\times 1]
      \quad \mbox{vector}             \nonumber  \\
\tilde\mathbf{w}_{k,\, m_1}
& = &
\left[
\begin{array}{ccc}
\tilde w_{k,\, m_1} & \ldots  & \tilde w_{k,\, \, m_1+L_p-{L_{hr}}}
\end{array}
\right]^T                             \nonumber  \\
&   & \mbox{ }
      [(L_p-L_{hr}+1)\times 1]
      \quad \mbox{vector}             \nonumber  \\
\tilde\mathbf{h}_k
& = &
\left[
\begin{array}{ccc}
\tilde h_{k,\, 0}        & \ldots  & \tilde h_{k,\, L_{hr}-1}
\end{array}
\right]^T                             \nonumber  \\
&   & \mbox{ }
      [L_{hr}\times 1]
      \quad \mbox{vector}             \nonumber  \\
\tilde\mathbf{s}_1
& = &
\left[
\begin{array}{ccc}
\tilde s_{1,\, L_{hr}-1} & \ldots  & \tilde s_{1,\, 0}\\
\vdots                   & \ldots  & \vdots\\
\tilde s_{1,\, L_p-1}    & \ldots  & \tilde s_{1,\, L_p-L_{hr}-2}
\end{array}
\right]                               \nonumber  \\
&   & \mbox{ }
      [(L_p-L_{hr}+1)\times L_{hr}]
      \quad \mbox{matrix}
\end{eqnarray}
%*******************************************************************************
where again $L_{hr}(> L_h)$ is the channel length assumed by the receiver.
The statement of the ML channel estimation is as follows: find
$\hat\mathbf{h}_k$ (the estimate of $\tilde\mathbf{h}_k$) such that:
%*******************************************************************************
\begin{eqnarray}
\label{Eq:Pap8_Eq20}
\left(
\tilde\mathbf{r}_{k,\, m_1} -
\tilde\mathbf{s}_1
\hat\mathbf{h}_k
\right)^H
\left(
\tilde\mathbf{r}_{k,\, m_1} -
\tilde\mathbf{s}_1
\hat\mathbf{h}_k
\right)
\end{eqnarray}
%*******************************************************************************
is minimized. Differentiating with respect to $\hat\mathbf{h}_k^*$ and
setting the result to zero yields \cite{Haykin_Adapt_96,Vasu_Book10}:
%*******************************************************************************
\begin{eqnarray}
\label{Eq:Pap8_Eq21}
\hat\mathbf{h}_k =
                 \left(
                 \tilde\mathbf{s}_1^H
                 \tilde\mathbf{s}_1
                 \right)^{-1}
                 \tilde\mathbf{s}_1^H
                 \tilde\mathbf{r}_{k,\, m_1}.
\end{eqnarray}
%*******************************************************************************
To see the effect of noise on the channel estimate in (\ref{Eq:Pap8_Eq21}),
consider
%*******************************************************************************
\begin{eqnarray}
\label{Eq:Pap8_Eq22}
\tilde\mathbf{u} =
                 \left(
                 \tilde\mathbf{s}_1^H
                 \tilde\mathbf{s}_1
                 \right)^{-1}
                 \tilde\mathbf{s}_1^H
                 \tilde\mathbf{w}_{k,\, m_1}.
\end{eqnarray}
%*******************************************************************************
When $m_0=L_h-1$, observe that
%*******************************************************************************
\begin{eqnarray}
\label{Eq:Pap8_Eq22_0}
\hat\mathbf{h}_k = \tilde\mathbf{h}_k + \tilde\mathbf{u}.
\end{eqnarray}
%*******************************************************************************
Since $\tilde s_{1,\, n}$ is a zero-mean random sequence with good
autocorrelation properties, it is reasonable to expect
%*******************************************************************************
\begin{eqnarray}
\label{Eq:Pap8_Eq23}
\tilde\mathbf{s}_1^H
\tilde\mathbf{s}_1 & = &  L_1
                         \sigma^2_s
                         \mathbf{I}_{L_{hr}}
                         \quad \mbox{for $L_p \gg L_{hr}$}  \nonumber  \\
\Rightarrow
\left(
\tilde\mathbf{s}_1^H
\tilde\mathbf{s}_1
\right)^{-1}       & = &  1/(L_1
                         \sigma^2_s)
                         \mathbf{I}_{L_{hr}}                \nonumber  \\
\Rightarrow
\tilde\mathbf{u}   & = &  1/(L_1
                         \sigma^2_s)
                         \tilde\mathbf{s}_1^H
                         \tilde\mathbf{w}_{k,\, m_1}
\end{eqnarray}
%*******************************************************************************
where $\sigma^2_s$ is defined in (\ref{Eq:Pap8_Eq4_2}), $L_1$ is defined in
(\ref{Eq:Pap8_Eq6_1}), and $\mathbf{I}_{L_{hr}}$ is an $L_{hr}\times L_{hr}$
identity matrix. It can be shown that
%*******************************************************************************
\begin{eqnarray}
\label{Eq:Pap8_Eq24}
 E
\left[
\tilde\mathbf{u}
\tilde\mathbf{u}^H
\right] = \frac{2\sigma^2_w}{L_1\sigma^2_s}
          \mathbf{I}_{L_{hr}}
        = \frac{\sigma^2_w L_d}{L_1}
          \mathbf{I}_{L_{hr}}
          \stackrel{\Delta}{=}  2
          \sigma^2_u \mathbf{I}_{L_{hr}}.
\end{eqnarray}
%*******************************************************************************
Therefore, the variance of the ML channel estimate ($\sigma^2_u$) tends to
zero as $L_1\rightarrow \infty$ and $L_d$ is kept fixed. Conversely, when
$L_d$ is increased keeping $L_1$ fixed, there is noise enhancement.
%*******************************************************************************
%Let $\tilde u_l$ denote an element of $\tilde\mathbf{u}$. Then
%*******************************************************************************
%\begin{eqnarray}
%\label{Eq:Pap8_Eq25}
%\tilde U_n = \sum_{l=0}^{L_{hr}-1}
%             \tilde u_l
%             \,
%             \mathrm{e}^{-\mathrm{j}\, 2\pi l n/L_d}
%             \quad \mbox{for $0 \le i \le L_d-1$}
%\end{eqnarray}
%*******************************************************************************
%is the $L_d$-point DFT of $\tilde\mathbf{u}$. It is clear that
%*******************************************************************************
%\begin{eqnarray}
%\label{Eq:Pap8_Eq26}
% E
%\left[
%\tilde U_n
%\tilde U_m^*
%\right] = \left(
%           2
%          \sigma^2_u
%          \right)
%          \sum_{l=0}^{L_{hr}-1}
%          \mathrm{e}^{-\mathrm{j}\, 2\pi l (n-m)/L_d}.
%\end{eqnarray}
%*******************************************************************************

At this point, it must be mentioned that in the absence of noise, the channel
estimate obtained from (\ref{Eq:Pap8_Eq21}) depends on the SoF estimate $m_0$
obtained from (\ref{Eq:Pap8_Eq16}). When $m_0=L_h-1$, the
channel estimate in the absence of noise would be:
%*******************************************************************************
\begin{eqnarray}
\label{Eq:Pap8_Eq27}
\hat\mathbf{h}_k =
\left[
\begin{array}{cccccc}
\tilde h_{k,\, 0} & \ldots & \tilde h_{k,\, L_h-1} & 0 & \ldots & 0
\end{array}
\right]^T
\end{eqnarray}
%*******************************************************************************
When $m_0=0$, the channel estimate (in the absence of noise) is :
%*******************************************************************************
\begin{eqnarray}
\label{Eq:Pap8_Eq29}
\hat\mathbf{h}_k =
\left[
\begin{array}{cccccc}
0 & \ldots & 0 & \tilde h_{k,\, 0} & \ldots & \tilde h_{k,\, L_h-1}
\end{array}
\right]^T.
\end{eqnarray}
%*******************************************************************************
Thus we get:
%*******************************************************************************
\begin{eqnarray}
\label{Eq:Pap8_Eq30}
L_{hr} = 2 L_h - 1.
\end{eqnarray}
%*******************************************************************************
Observe that the channel estimation matrix $\tilde\mathbf{s}_1$ in
(\ref{Eq:Pap8_Eq19}) remains the same, independent of $m_0$. Therefore, the
pseudoinverse of $\tilde\mathbf{s}_1$ given in (\ref{Eq:Pap8_Eq21}) can be
precomputed and stored in the receiver.
%*******************************************************************************
%\begin{figure}[tbh]
%\centering
%\input{dmt17_estch_p512_o512_snr0.pstex_t}
%\caption{Magnitude response of the channel at 0 dB SNR per bit after fine
%         frequency offset compensation, $L_p=512$, $L_d=512$.}
%\label{Fig:DMT17_Estch_P512_O512_SNR0}
%\end{figure}
%*******************************************************************************
%*******************************************************************************
\begin{figure}[tbh]
\centering
\input{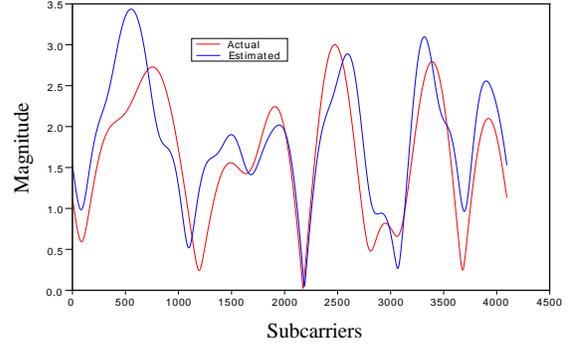}
\caption{Magnitude response of the channel at 0 dB SNR per bit after fine
         frequency offset compensation, $L_p=512$, $L_d=4096$.}
\label{Fig:DMT17_Estch_P512_O4096_SNR0}
\end{figure}
%*******************************************************************************
%*******************************************************************************
\begin{figure}[tbh]
\centering
\input{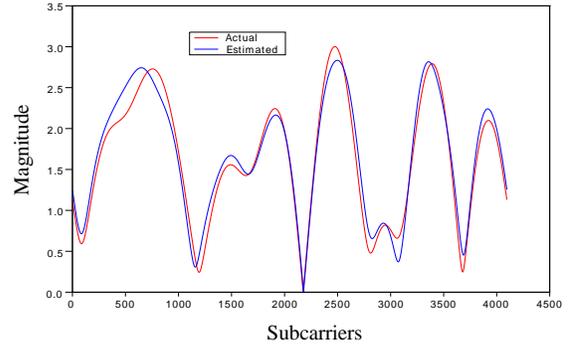}
\caption{Magnitude response of the channel at 10 dB SNR per bit after fine
         frequency offset compensation, $L_p=512$, $L_d=4096$.}
\label{Fig:DMT17_Estch_P512_O4096_SNR10}
\end{figure}
%*******************************************************************************
The magnitude response of the channel at various SNRs are shown in
%Figures~\ref{Fig:DMT17_Estch_P512_O512_SNR0},
Figures~\ref{Fig:DMT17_Estch_P512_O4096_SNR0} and
\ref{Fig:DMT17_Estch_P512_O4096_SNR10}.

%*******************************************************************************
\subsection{Fine Frequency Offset Estimation}
\label{SSec:Fine_FOFF_Est}
\begin{figure}[tbh]
\centering
\input{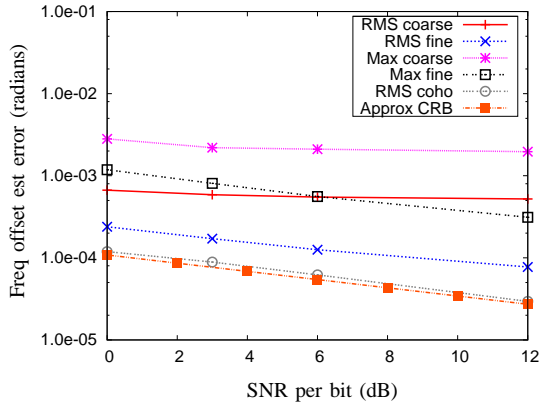}
\caption{RMS and maximum frequency offset estimation error for $L_p=512$.}
\label{Fig:DMT17_FOFF}
\end{figure}

%*******************************************************************************
For the purpose of fine frequency offset estimation, we propose to use
(\ref{Eq:Pap8_Eq14}) with $\tilde y_{k,\, n}$ replaced by $\hat y_{k,\, n}$ as
given in (\ref{Eq:Pap8_Eq15}). Moreover, since the initial estimate of the
frequency offset $(\hat\omega_k)$ is already available, (\ref{Eq:Pap8_Eq14})
must be modified as follows:
%*******************************************************************************
\begin{eqnarray}
\label{Eq:Pap8_Eq31}
\max_{m,\,\hat\omega_{k,\, f}}
                       \left|
                       \sum_{i=0}^{L_2-1}
                       \tilde r_{m+i}\,\hat y_{k,\, i}^*
                       \,
                       \mathrm{e}^{-\mathrm{j}\,
                       \left(
                       \hat\omega_k+\hat\omega_{k,\, f}
                       \right) i}
                       \right|
\end{eqnarray}
%*******************************************************************************
where
%*******************************************************************************
\begin{eqnarray}
\label{Eq:Pap8_Eq32}
L_2 & = & L_{hr} + L_p - 1                        \nonumber  \\
0   & \le
        & m
      \le   L_{hr} - 1.
\end{eqnarray}
%*******************************************************************************
Observe that the span of $\hat y_{k,\, i}$ is $L_2$. The fine frequency offset
estimate ($\hat\omega_{k,\, f}$) is obtained by dividing the interval
$[\hat\omega_k-0.005,\, \hat\omega_k+0.005]$ radian into $B_2$ frequency bins
\cite{Vasu_SIVP10}. The reason for choosing 0.005 radian can be traced to
Figure~\ref{Fig:DMT17_FOFF}. We find that the maximum error in the coarse
estimate of the frequency offset is approximately 0.004 radian over
$10^4$ frames. Thus the probability that the maximum error exceeds 0.005
radian is less than $10^{-4}$.

In Figure~\ref{Fig:DMT17_FOFF}, the coarse frequency offset estimate is
obtained from (\ref{Eq:Pap8_Eq16}), fine frequency offset estimate from
(\ref{Eq:Pap8_Eq31}), coherent frequency offset estimate (``RMS coho'') from
(\ref{Eq:Ap_Eq4_1}) and the approximate Cram{\' e}r-Rao bound from
(\ref{Eq:Ap_Eq15}).
%*******************************************************************************
%*******************************************************************************

\begin{figure}[tbh]
\centering
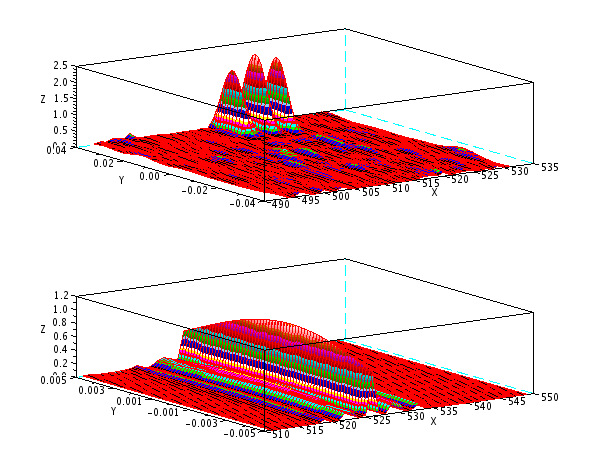
\caption{SoF detection, coarse and fine frequency offset estimation, for
         $L_p=512$, SNR per bit equal to 0 dB, $B_1=B_2=64$.
         \copyright\, 2013 IEEE. Reprinted, with permission, from
         \cite{Vasu13}.}
\label{Fig:DMT9_SOF_0db}
\end{figure}

Figure~\ref{Fig:DMT9_SOF_0db} gives the results for SoF detection, coarse and
fine frequency offset estimation,
for one particular frame at 0 dB SNR per bit, with $B_1=B_2=64$.
%*******************************************************************************
%*******************************************************************************
The advantage of the two-stage approach (coarse and fine) for frequency offset
estimation \cite{Vasu_SIVP10} is illustrated in
Table~\ref{Tbl:Foff_Complexity}. The complexity
of the two-stage approach is $B_1+B_2=128$ frequency bins. The resolution of
the two-stage approach is $2\times 0.005/B_2=0.00015625$ radian. For obtaining
the same resolution, the single stage approach will require
$2\times 0.04/0.00015625=512$ frequency
bins. Therefore, the two-stage approach is four times more efficient than the
single stage approach.
%*******************************************************************************
\begin{table}[tbh]
\centering
\caption{Complexity comparison between the two-stage and single stage approach
         of frequency offset estimation.}
\input{foff_complexity.pstex_t}
\label{Tbl:Foff_Complexity}
\end{table}
%*******************************************************************************

At this point, a note on the implementation of the SoF and frequency offset
estimation algorithm is in order. Observe that a 2-D search over both
frequency and time is required and there is a large scope for parallel
processing. Hence, this algorithm is well suited for hardware implementation.

%*******************************************************************************
\subsection{Noise Variance Estimation}
\label{SSec:Noise_Var_Est}
%*******************************************************************************
It is necessary to estimate the noise variance for the purpose of turbo
decoding \cite{Vasu_Book10}. After the channel has been estimated using
(\ref{Eq:Pap8_Eq21}), the noise variance is estimated as follows:
%*******************************************************************************
\begin{eqnarray}
\label{Eq:Pap8_Eq33}
\hat
\sigma^2_w = \frac{1}{2 L_1}
\left(
\tilde\mathbf{r}_{k,\, m_1} -
\tilde\mathbf{s}_1
\hat\mathbf{h}_k
\right)^H
\left(
\tilde\mathbf{r}_{k,\, m_1} -
\tilde\mathbf{s}_1
\hat\mathbf{h}_k
\right)
\end{eqnarray}
%*******************************************************************************
where $\tilde\mathbf{s}_1$ is defined in (\ref{Eq:Pap8_Eq19}) and $L_1$ is
defined in (\ref{Eq:Pap8_Eq6_1}).
%*******************************************************************************
\subsection{Turbo Decoding}
\label{SSec:Turbo_Dec}
\begin{figure}[tbh]
\centering
\input{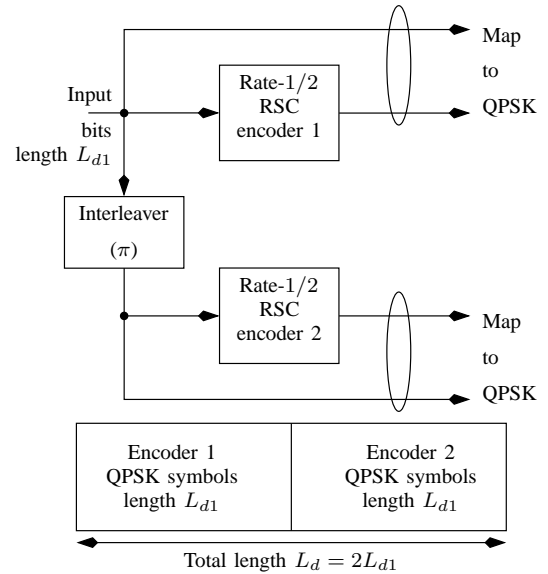}
\caption{Encoder block diagram.}
\label{Fig:Turbo_Enc}
\end{figure}

\begin{figure}[tbh]
\centering
\input{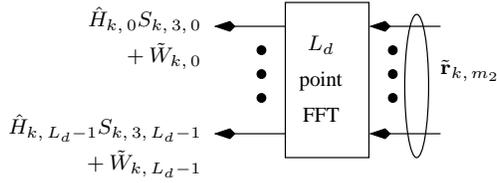}
\caption{OFDM receiver after synchronization.}
\label{Fig:OFDM_Rx}
\end{figure}

%*******************************************************************************
The encoder block diagram is shown in Figure~\ref{Fig:Turbo_Enc}. The overall
rate of the encoder is $1/2$, since $L_{d1}$ data bits generate $2L_{d1}$
coded QPSK symbols.
%*******************************************************************************
%*******************************************************************************
The generating matrix for each of the constituent encoders is given by:
%*******************************************************************************
\begin{eqnarray}
\label{Eq:Pap8_Eq34}
\mathbf{G}(D) =
\left[
\begin{array}{cc}
 1 & \frac{\displaystyle 1+D^2}{\displaystyle 1+D+D^2}
\end{array}
\right].
\end{eqnarray}
%*******************************************************************************
Let
%*******************************************************************************
\begin{eqnarray}
\label{Eq:Pap8_Eq35}
m_2 = m_1 + L_p
\end{eqnarray}
%*******************************************************************************
where $m_1$ is defined in (\ref{Eq:Pap8_Eq18_0}). Define
%*******************************************************************************
\begin{eqnarray}
\label{Eq:Pap8_Eq36}
\tilde\mathbf{r}_{k,\, m_2} =
\left[
\begin{array}{ccc}
\tilde r_{k,\, m_2} & \ldots & \tilde r_{k,\, m_2+L_d-1}
\end{array}
\right]
\end{eqnarray}
%*******************************************************************************
as the data part of the received signal for the $k^{th}$ frame.
After SoF detection, frequency offset compensation and channel estimation,
the receiver block diagram is depicted in Figure~\ref{Fig:OFDM_Rx}.
%*******************************************************************************
%*******************************************************************************
The output of the FFT can be written as (for $0 \le i \le L_d-1$):
%*******************************************************************************
\begin{eqnarray}
\label{Eq:Pap8_Eq36_1}
\tilde R_{k,\, i} = \hat H_{k,\, i} S_{k,\, 3,\, i} +
                    \tilde W_{k,\, i}.
\end{eqnarray}
%*******************************************************************************
Note that $\hat H_{k,\, i}$ and $\tilde W_{k,\, i}$ in
Figure~\ref{Fig:OFDM_Rx} are the $L_d$-point
DFT of the estimated channel $\hat\mathbf{h}_k$ in (\ref{Eq:Pap8_Eq21}) and
$\tilde w_{k,\, n}$
in (\ref{Eq:Pap8_Eq5}) respectively, taken over the time interval
specified in (\ref{Eq:Pap8_Eq36}), and $S_{k,\, 3,\, i}$ denotes the
data symbols for the $k^{th}$ frame, for $0 \le i \le L_d-1$.

The variance of $\tilde W_{k,\, i}$ is
%*******************************************************************************
\begin{eqnarray}
\label{Eq:Pap8_Eq37}
\frac{1}{2}
 E
\left[
\left|
\tilde W_{k,\, i}
\right|^2
\right] = L_d \sigma^2_w
\end{eqnarray}
%*******************************************************************************
and the variance of $\hat H_{k,\, i}$ is (assuming perfect channel estimates,
that is $\hat H_{k,\, i}=\tilde H_{k,\, i}$):
%*******************************************************************************
\begin{eqnarray}
\label{Eq:Pap8_Eq38}
\frac{1}{2}
 E
\left[
\left|
\tilde H_{k,\, i}
\right|^2
\right] = L_h \sigma^2_f.
\end{eqnarray}
%*******************************************************************************
Note that due to multiplication by the channel DFT ($\hat H_{k,\, i}$) in
(\ref{Eq:Pap8_Eq36_1}), the data and parity bits of the QPSK symbol cannot be
separated, and the BCJR
algorithm is slightly different from the one given in \cite{Vasu_Book10}.
This is explained below. Observe also that dividing (\ref{Eq:Pap8_Eq36_1})
by $\tilde H_{k,\, i}$ results in interference
($\tilde W_{k,\, i}/\hat H_{k,\, i}$) having a complex ratio
distribution \cite{Baxley10,Torres10}, which is undesirable.

Corresponding to the transition from state $m$ to state $n$,
at decoder 1, for the $k^{th}$ frame, at time $i$ define
(for $0 \le i \le L_{d1}-1$, $L_{d1}$ is defined in
Figure~\ref{Fig:Turbo_Enc}):
%*******************************************************************************
\begin{eqnarray}
\label{Eq:Turbo_Eq10}
\gamma_{1,\, k,\, i,\, m,\, n} =
                            \exp
                            \left[-
                            \frac{
                            \left(
                            \tilde R_{k,\, i}-
                            \hat H_{k,\, i}
                             S_{m,\, n}
                            \right)^2}
                                 {2L_d\hat\sigma_w^2}
                                              \right]
\end{eqnarray}
%*******************************************************************************
where $S_{m,\, n}$ denotes the QPSK symbol corresponding to the transition
from state $m$ to state $n$ in the trellis. We assume that the data bit maps
to the real part and the parity bit maps to the imaginary part of the QPSK
symbol. We also assume that bit 0 maps to $+1$ and bit 1 maps to $-1$. Observe
that $\hat\sigma^2_w$ is the estimate of $\sigma^2_w$ obtained from
(\ref{Eq:Pap8_Eq33}). Similarly, for the
transition from state $m$ to state $n$,
at decoder 2, for the $k^{th}$ frame, at time $i$ define
(for $0 \le i \le L_{d1}-1$):
%*******************************************************************************
\begin{eqnarray}
\label{Eq:Turbo_Eq10_1}
\gamma_{2,\, k,\, i,\, m,\, n} =
                            \exp
                            \left[-
                            \frac{
                            \left(
                            \tilde R_{k,\, L_{d1}+i}-
                            \hat H_{k,\, L_{d1}+i}
                             S_{m,\, n}
                            \right)^2}
                                 {2L_d\hat\sigma_w^2}
                                              \right]
\end{eqnarray}
%*******************************************************************************
Let $\mathscr{S}$ denote the number of states in
the encoder trellis. Let $\mathscr{D}_n$ denote the set of states that diverge
from state $n$. For example
%*******************************************************************************
\begin{eqnarray}
\label{Eq:Turbo_Eq8_2}
\mathscr{D}_0 = \{0,\, 3\}
\end{eqnarray}
%*******************************************************************************
implies that states 0 and 3 can be reached from state 0.
Similarly, let $\mathscr{C}_n$ denote the set of states that converge
to state $n$. Let $\alpha_{i,\, n}$ denote the alpha value at time $i$
($0 \le i \le L_{d1}-2$) at state $n$ ($0 \le n \le \mathscr{S}-1$).

Then the alpha values for decoder 1 can be recursively computed as follows
(forward recursion):
%*******************************************************************************
\begin{eqnarray}
\label{Eq:Turbo_Eq9}
\alpha_{i+1,\, n}' & = & \sum_{m \in \mathscr{C}_n}
                         \alpha_{i,\, m}
                         \gamma_{1,\, k,\, i,\, m,\, n}
                          P
                         \left(
                          S_{b,\, i,\, m,\, n}
                         \right)                      \nonumber  \\
\alpha_{0,\, n}    & = &  1
                         \qquad
                         \mbox{for $0 \le n \le \mathscr{S}-1$}
                                                      \nonumber  \\
\alpha_{i+1,\, n}  & = & \alpha_{i+1,\, n}'\Big/
                         \left(
                         \sum_{n=0}^{\mathscr{S}-1}
                         \alpha_{i+1,\, n}'
                         \right)
\end{eqnarray}
%*******************************************************************************
where
%*******************************************************************************
\begin{eqnarray}
\label{Eq:Turbo_Eq9_1}
P(S_{b,\, i,\, m,\, n}) =
\left
\{
\begin{array}{ll}
F_{2,\, i+} & \mbox{if $S_{b,\, i,\, m,\, n}=+1$}\\
F_{2,\, i-} & \mbox{if $S_{b,\, i,\, m,\, n}=-1$}\\
\end{array}
\right.
\end{eqnarray}
%*******************************************************************************
denotes the {\it a priori\/} probability of the
systematic bit corresponding to the transition from state $m$ to state $n$,
at decoder 1, at time $i$, obtained from the $2^{nd}$ decoder at time $l$,
after deinterleaving (that is, $i=\pi^{-1}(l)$ for some
$0 \le l \le L_{d1}-1$).
The terms $F_{2,\, i+}$ and $F_{2,\, i-}$ are defined similar to
(\ref{Eq:Turbo_Eq12_2}) given below.
The normalization step in the last equation of
(\ref{Eq:Turbo_Eq9}) is done to prevent numerical instabilities
\cite{Vasu_Book10,Singer04}.

Similarly, let $\beta_{i,\, n}$ denote the
beta values at time $i$ ($1 \le i \le L_{d1}-1$) at state $n$
($0 \le n \le \mathscr{S}-1$).
Then the recursion for beta (backward recursion) at decoder 1
can be written as:
%*******************************************************************************
\begin{eqnarray}
\label{Eq:Turbo_Eq11}
\beta_{i,\, n}' & = & \sum_{m \in \mathscr{D}_n}
                      \beta_{i+1,\, m}
                      \gamma_{1,\, k,\, i,\, n,\, m}
                       P
                      \left(
                       S_{b,\, i,\, n,\, m}
                      \right)                                 \nonumber  \\
\beta_{L_{d1},\, n}
                & = &  1
                      \qquad
                      \mbox{for $0 \le n \le \mathscr{S}-1$}  \nonumber  \\
\beta_{i,\, n}  & = & \beta_{i,\, n}'\Big/
                      \left(
                      \sum_{n=0}^{\mathscr{S}-1}
                      \beta_{i,\, n}'
                      \right).
\end{eqnarray}
%*******************************************************************************
Once again, the normalization step in the last equation of
(\ref{Eq:Turbo_Eq11}) is done to prevent numerical instabilities.

Let $\rho^+(n)$ denote the state that is reached from
state $n$ when the input symbol is $+1$. Similarly let $\rho^-(n)$ denote
the state that can be reached from state $n$ when the input symbol is $-1$.
Then (for $0 \le i \le L_{d1}-1$)
%*******************************************************************************
\begin{eqnarray}
\label{Eq:Turbo_Eq12}
G_{1,\, i+}
       & = & \sum_{n=0}^{\mathscr{S}-1}
             \alpha_{i,\, n}
             \gamma_{1,\, k,\, i,\, n,\, \rho^+(n)}
             \beta_{i+1,\, \rho^+(n)}                      \nonumber  \\
G_{1,\, i-}
       & = & \sum_{n=0}^{\mathscr{S}-1}
             \alpha_{i,\, n}
             \gamma_{1,\, k,\, i,\, n,\, \rho^-(n)}
             \beta_{i+1,\, \rho^-(n)}.
\end{eqnarray}
%*******************************************************************************
Finally, the extrinsic information that is to be fed as {\it a priori\/}
probabilities to the second decoder after interleaving, is computed as:
%*******************************************************************************
\begin{eqnarray}
\label{Eq:Turbo_Eq12_2}
F_{1,\, i+}
& = & G_{1,\, i+}/
     (G_{1,\, i+} + G_{1,\, i-}) \nonumber  \\
F_{1,\, i-}
& = & G_{1,\, i-}/
     (G_{1,\, i+} + G_{1,\, i-})
\end{eqnarray}
%*******************************************************************************
Equations (\ref{Eq:Turbo_Eq9}), (\ref{Eq:Turbo_Eq11}),
(\ref{Eq:Turbo_Eq12}) and (\ref{Eq:Turbo_Eq12_2}) constitute the MAP
recursions for the first decoder. The MAP recursions for the second decoder
are similar.

After a few iterations, (one iteration involves both decoder 1 and 2) the
final {\it a posteriori\/} probabilities of $i^{th}$ bit of the $k^{th}$
frame at the output of decoder 1 is given by:
%*******************************************************************************
\begin{eqnarray}
\label{Eq:Turbo_Eq13}
H_{1,\, i+}
        & = & \sum_{n=0}^{\mathscr{S}-1}
              \alpha_{i,\, n}
              \gamma_{1,\, k,\, i,\, n,\, \rho^+(n)}
               F_{2,\, i+}
              \,
              \beta_{i+1,\, \rho^+(n)}                      \nonumber  \\
H_{1,\, i-}
        & = & \sum_{n=0}^{\mathscr{S}-1}
              \alpha_{i,\, n}
              \gamma_{1,\, k,\, i,\, n,\, \rho^-(n)}
               F_{2,\, i-}
              \,
              \beta_{i+1,\, \rho^-(n)}.                     \nonumber  \\
\end{eqnarray}
%*******************************************************************************
followed by
%*******************************************************************************
\begin{eqnarray}
\label{Eq:Turbo_Eq14}
P
\left(
S_{b,\, k,\, i}= +1|\mathbf{r}_{k,\, m2}
\right) & = &  H_{1,\, i+}/
              (H_{1,\, i+} + H_{1,\, i-})               \nonumber  \\
P
\left(
S_{b,\, k,\, i}= -1|\mathbf{r}_{k,\, m2}
\right) & = &  H_{1,\, i-}/
              (H_{1,\, i+} + H_{1,\, i-}).              \nonumber  \\
\end{eqnarray}
%*******************************************************************************
When puncturing is used to increase the overall rate, e.g. if the QPSK symbol
occurring at odd instants of time in both encoders are not transmitted, then
the corresponding gamma values in (\ref{Eq:Turbo_Eq10}) and
(\ref{Eq:Turbo_Eq10_1}) are set to unity. For the even time instants, the
corresponding gamma values are computed according to (\ref{Eq:Turbo_Eq10}) and
(\ref{Eq:Turbo_Eq10_1}).
%*******************************************************************************
\subsection{Robust Turbo Decoding}
\label{SSec:Robust_Turbo_Dec}
%*******************************************************************************
At high SNR, the term in the exponent ($b$ is the exponent of $\mathrm{e}^b$)
of (\ref{Eq:Turbo_Eq10}) and (\ref{Eq:Turbo_Eq10_1}) becomes very large
(typically $b>100$) and it becomes unfeasible for the DSP processor or even a
computer to calculate the gammas. We propose to solve this problem by
normalizing the exponents. Observe that the exponents are real-valued and
negative. Let $b_{1,\, j,\, i}$ denote an exponent at decoder 1 due to the
$j^{th}$ symbol in the constellation ($1\le j\le 4$ for QPSK) at time $i$.
Let
%*******************************************************************************
\begin{eqnarray}
\label{Eq:Turbo_Eq15}
\mathbf{b}_1 =
\left[
\begin{array}{ccc}
b_{1,\, 1,\, 0} & \ldots & b_{1,\, 1,\, L_{d1}-1}\\
\vdots          & \vdots & \vdots                \\
b_{1,\, 4,\, 0} & \ldots & b_{1,\, 4,\, L_{d1}-1}
\end{array}
\right]
\end{eqnarray}
%*******************************************************************************
denote the matrix of exponents for decoder 1. Let $b_{1,\, \mathrm{max},\, i}$
denote the maximum exponent at time $i$, that is
%*******************************************************************************
\begin{eqnarray}
\label{Eq:Turbo_Eq15_1}
b_{1,\,\mathrm{max},\, i} =
\max
\left[
\begin{array}{c}
b_{1,\, 1,\, i}\\
\vdots         \\
b_{1,\, 4,\, i}
\end{array}
\right].
\end{eqnarray}
%*******************************************************************************
Let
%*******************************************************************************
\begin{eqnarray}
\label{Eq:Turbo_Eq15_2}
\mathbf{b}_{1,\,\mathrm{max}} =
\left[
\begin{array}{ccc}
b_{1,\, \mathrm{max},\, 0} & \ldots & b_{1,\,\mathrm{max},\, L_{d1}-1}
\end{array}
\right]
\end{eqnarray}
%*******************************************************************************
denote the vector containing the maximum exponents. Compute:
%*******************************************************************************
\begin{eqnarray}
\label{Eq:Turbo_Eq16}
\mathbf{b}_1' = \mathbf{b}_1 -
\left[
\begin{array}{c}
\mathbf{b}_{1,\,\mathrm{max}}\\
\vdots\\
\mathbf{b}_{1,\,\mathrm{max}}
\end{array}
\right].
\end{eqnarray}
%*******************************************************************************
Note that in (\ref{Eq:Turbo_Eq16}), the vector $\mathbf{b}_{1,\,\mathrm{max}}$
has to be repeated as many times as the number of symbols in the constellation.

If any element of $\mathbf{b}_1'$ is less than say, $-30$, then set it to
$-30$. Thus we get a normalized exponent vector
$\mathbf{b}_{1,\,\mathrm{norm}}$,
whose elements lie in the range $[0,\, -30]$. It has been found from
simulations that normalizing the exponents does not lead to any degradation
in BER performance, on the contrary, it increases the operating SNR range of
the turbo receiver. In practice, we could divide the range $[0,\, -30]$ into
a large number (e.g. 3000) of levels and the exponentials ($\mathrm{e}^b$)
could be precomputed and stored in the DSP processor, and need not be
computed in real-time. The choice of the minimum exponent (e.g. $-30$),
would depend on the precision of the DSP processor or the computer.
%*******************************************************************************
\subsection{Data Interleaving}
\label{SSec:Data_Iv}
%*******************************************************************************
Assuming ideal channel estimates, the autocorrelation of the channel DFT at
the receiver is:
%*******************************************************************************
\begin{eqnarray}
\label{Eq:Pap8_Eq38_1}
\frac{1}{2}
 E
\left[
\tilde H_{k,\, i}
\tilde H_{k,\, j}^*
\right] = \sigma^2_f
          \sum_{n=0}^{L_h-1}
          \mathrm{e}^{-\mathrm{j}\, 2\pi n (i-j)/L_d}.
\end{eqnarray}
%*******************************************************************************
It has been found from simulations that the performance of the turbo decoder
gets adversely affected due to the correlation in $\tilde H_{k,\, i}$. To
overcome this problem, we interleave the data before the IFFT operation
at the transmitter and deinterleave the data after the FFT operation at the
receiver, before turbo decoding. This process essentially removes any
correlation in $\tilde H_{k,\, i}$ \cite{Fischer01}.
%*******************************************************************************
\subsection{Enhanced Frame Structure}
\label{SSec:Enhanced_Frame_Struct}
%*******************************************************************************
The accuracy of the frequency offset estimate depends on the length of the
preamble $L_p$. Increasing the number of frequency bins $B_1$ and $B_2$ in
Figure~\ref{Fig:DMT9_SOF_0db}, for a given $L_p$, does not improve the
accuracy. From Figure~\ref{Fig:DMT17_FOFF} it can be seen that the RMS value
of the fine frequency offset estimation error is about $2\times 10^{-4}$,
at an SNR per bit equal to 8 dB. The subcarrier spacing with data length
$L_d=4096$ is
equal to $2\pi/4096=1.534\times 10^{-3}$ radians. Therefore, the residual
frequency offset is $0.0002\times 100/0.001534=13\%$ of the subcarrier spacing,
which is quite high and causes severe intercarrier interference (ICI). Note
that the RMS frequency offset estimation error can be reduced by increasing
the preamble length ($L_p$), keeping the data length ($L_d$) fixed, which in
turn reduces the throughput given by:
%*******************************************************************************
\begin{eqnarray}
\label{Eq:Pap8_Eq38_2}
\mathscr{T} = \frac{L_{d1}}{L_p+L_{cp}+L_d}.
\end{eqnarray}
%*******************************************************************************
Note that for a rate-$1/2$ turbo code $L_d=2L_{d1}$, whereas for a rate-1
turbo code, $L_d=L_{d1}$.
This motivates us to
look for an alternate frame structure which not only solves the frequency
offset estimation problem, but also maintains the throughput at a reasonable
value.
%*******************************************************************************
\begin{figure}[tbh]
\centering
\input{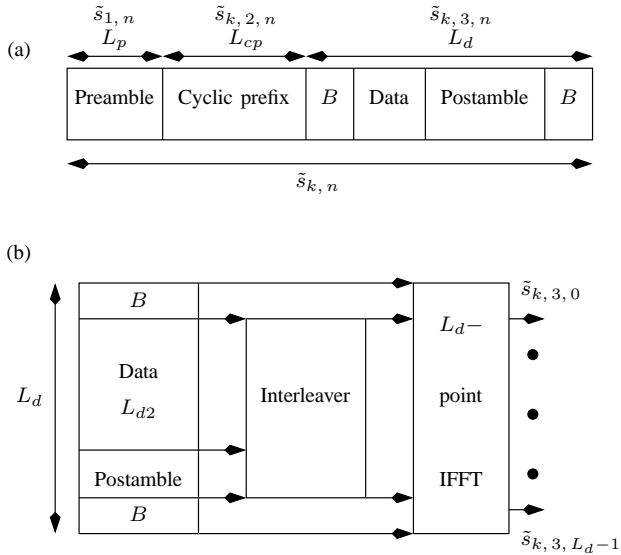}
\caption{(a) Enhanced frame structure. (b) Processing of the data part at the
         transmitter.}
\label{Fig:Enh_Frame}
\end{figure}
%*******************************************************************************

Consider the frame in Figure~\ref{Fig:Enh_Frame}(a). In addition to the
preamble, prefix and data, it contains ``buffer'' (dummy) symbols of length $B$
and postamble of length $L_o$, all drawn from the QPSK constellation. In
Figure~\ref{Fig:Enh_Frame}(b) we illustrate the processing of $L_d$ symbols
at the transmitter.
Observe that only the data and postamble symbols are interleaved before the
IFFT operation. After interleaving, the postamble gets randomly dispersed
between the data symbols. The buffer symbols are sent directly to the IFFT,
without interleaving. The preamble and the cyclic prefix continue to be
processed according to Figure~\ref{Fig:Frame} and (\ref{Eq:Pap8_Eq4_1}). We
now explain the reason behind using this frame structure. In what follows,
we assume that the SoF has been detected, fine frequency offset
correction has been performed and the channel has been estimated.

\begin{figure}[tbh]
\centering
\input{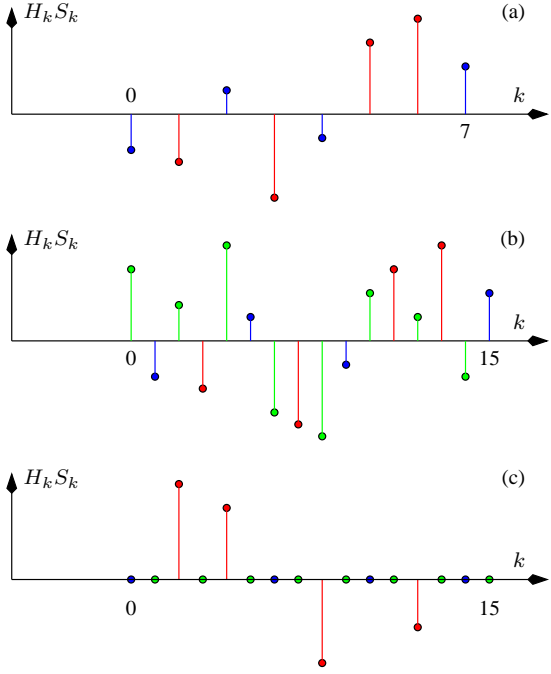}
\caption{Obtaining the postamble matched filter for $L_d=8$. Buffer symbols
         are not shown. The frequency offset ($\pi/L_d$) is half the
         subcarrier spacing
         ($2\pi/L_d$). $H_k$ and $S_k$ are assumed to be real-valued. Noise is
         absent. (a) Output of the $L_d$-point FFT in the absence of frequency
         offset. The red lines represent postamble and the blue lines
         represent data symbols. (b) Output of the $2L_d$-point FFT
         in the presence of frequency offset. Observe that the red and blue
         lines have shifted to the right by $\pi/L_d$. Green lines denote the
         output of the $L_d$-point FFT in the presence of frequency offset.
         (c) The postamble matched filter.}
\label{Fig:Post_MF}
\end{figure}

We proceed by making the following observations:
%*******************************************************************************
\begin{enumerate}
 \item Modulation in the time domain results in a shift in the frequency
       domain. Therefore, any residual frequency offset after fine
       frequency offset correction, results in a frequency shift at the
       output of the FFT operation at the receiver. Moreover, due to the
       presence of a cyclic prefix, the frequency shift is circular.
       Therefore, without the buffer symbols, there is a possibility that
       the first data symbol would be circularly shifted to the last
       data symbol or vice versa. This
       explains the use of buffer symbols at both ends in
       Figure~\ref{Fig:Enh_Frame}. In order to compute the number of buffer
       symbols ($B$), we have to know the maximum residual
       frequency offset, after fine frequency offset correction. Referring
       to Figure~\ref{Fig:DMT17_FOFF}, we find that the maximum error in
       fine frequency offset estimation at 0 dB SNR per bit is about
       $\pm 2\times 10^{-3}$ radians. With $L_d=4096$, the subcarrier spacing
       is $2\pi/4096=1.534\times 10^{-3}$ radians. Hence, the residual
       frequency error would result in a shift of $\pm 2/1.534=\pm 1.3$
       subcarrier spacings. Therefore, while $B=2$ would suffice, we have
       taken $B=4$, to be on the safe side.
 \item Since the frequency shift is not an integer multiple of the subcarrier
       spacing, we need to interpolate in between the subcarriers, to
       accurately estimate the shift. Interpolation can be achieved by
       zero-padding the data before the FFT operation. Thus we get a
       $2L_d-$point FFT corresponding to an interpolation factor of 2 and
       so on. Other methods of interpolation between subcarriers is
       discussed in \cite{Darshan13}.
 \item After the FFT operation, postamble matched filtering has to be
       done, since the postamble and $\hat H_k\approx \tilde H_k$ (in
       (\ref{Eq:Pap8_Eq36_1})) are available. The procedure for constructing
       the postamble matched filter is illustrated in
       Figure~\ref{Fig:Post_MF}. From simulations, it has been
       found that a postamble length $L_o=128$ results in false peaks at
       the postamble matched filter output at 0 dB SNR per bit. Therefore we
       have taken $L_o=256$. With these calculations, the length of the
       data works out as $L_{d2}=L_d-2B-L_o=4096-8-256=3832$ QPSK symbols.
       The throughput of the proposed system (with rate-1 turbo code) is
%       $L_{d2}/(L_p+L_d)=3832/(512+4096)=83.2\%$.
%*******************************************************************************
\begin{eqnarray}
\label{Eq:Pap8_Eq38_2_1}
\mathscr{T} & = & \frac{L_{d2}}{L_p+L_{cp}+L_d}    \nonumber  \\
            & = & \frac{3832}{512+18+4096}         \nonumber  \\
            & = & 82.84\%.
\end{eqnarray}
%*******************************************************************************
       The throughput comparison of various
       frame structures is summarized in Table~\ref{Tbl:Throughput_Comp}.
\end{enumerate}
%*******************************************************************************
%*******************************************************************************
\begin{table}[tbh]
\centering
\caption{Throughput comparison of various frame structures with
         $L_p=L_{d1}=512$, $L_{d2}=3832$, $L_{cp}=18$.}
\input{throughput_comp.pstex_t}
\label{Tbl:Throughput_Comp}
\end{table}
%*******************************************************************************
%*******************************************************************************
%*******************************************************************************
%*******************************************************************************
\subsection{Receiver Diversity}
\label{SSec:Rx_Div}
%*******************************************************************************
In the presence of receiver diversity, the signal in each diversity
arm ($l$) can be expressed as (see (\ref{Eq:Pap8_Eq5})):
%*******************************************************************************
\begin{eqnarray}
\label{Eq:Pap8_Eq38_3}
\tilde r_{k,\, n,\, l} & = & \left(
                             \tilde s_{k,\, n} \star \tilde h_{k,\, n,\, l}
                             \right)\,
                             \mathrm{e}^{\,\mathrm{j}
                            (\omega_k n+\theta_{k,\, l})} +
                             \tilde w_{k,\, n,\, l}            \nonumber  \\
                       & = & \tilde y_{k,\, n,\, l}
                             \mathrm{e}^{\,\mathrm{j}
                            (\omega_k n+\theta_{k,\, l})} +
                             \tilde w_{k,\, n,\, l}
\end{eqnarray}
%*******************************************************************************
for $1 \le l \le N$.
The frequency offset is assumed to be identical for all the diversity arms,
whereas the carrier phase and noise are assumed to be independent. The noise
variance is same for all the diversity arms.
Two extreme scenarios are considered  in the simulations (a) identical channel
and (b) independent channel in each diversity arm.
The output of the FFT can be written as (for $0 \le i \le L_d-1$):
%*******************************************************************************
\begin{eqnarray}
\label{Eq:Pap8_Eq38_3_1}
\tilde R_{k,\, i,\, l} = \hat H_{k,\, i,\, l} S_{k,\, 3,\, i} +
                         \tilde W_{k,\, i,\, l}
\end{eqnarray}
%*******************************************************************************
for $1 \le l \le N$ diversity arms.
The notation in (\ref{Eq:Pap8_Eq38_3_1}) is self explanatory and is based on
(\ref{Eq:Pap8_Eq36_1}).

In the turbo decoding operation, (for decoder 1, $N$ diversity arms,
rate-$1/2$ turbo code, the enhanced frame structure in
Figure~\ref{Fig:Enh_Frame} and $0 \le i \le L_{d2}/2 -1$), we have from
(\ref{Eq:Turbo_Eq10}):
%*******************************************************************************
\begin{eqnarray}
\label{Eq:Turbo_Eq38_4}
\gamma_{1,\, k,\, i,\, m,\, n} =
\prod_{l=1}^{N}
\gamma_{1,\, k,\, i,\, m,\, n,\, l}
\end{eqnarray}
%*******************************************************************************
where
%*******************************************************************************
\begin{eqnarray}
\label{Eq:Turbo_Eq38_5}
\gamma_{1,\, k,\, i,\, m,\, n,\, l} =
                            \exp
                            \left[-
                            \frac{
                            \left(
                            \tilde R_{k,\, i,\, l}-
                            \hat H_{k,\, i,\, l}
                             S_{m,\, n}
                            \right)^2}{2L_d\hat\sigma_w^2}
                            \right]
\end{eqnarray}
%*******************************************************************************
where $\hat\sigma_w^2$ is the average estimate of the noise variance over
all the diversity arms.
Similarly at decoder 2, for $0 \le i \le L_{d2}/2-1$,
we have from (\ref{Eq:Turbo_Eq10_1}):
%*******************************************************************************
\begin{eqnarray}
\label{Eq:Turbo_Eq38_6}
\gamma_{2,\, k,\, i,\, m,\, n} =
\prod_{l=1}^{N}
\gamma_{2,\, k,\, i,\, m,\, n,\, l}
\end{eqnarray}
%*******************************************************************************
where
%*******************************************************************************
\begin{eqnarray}
\label{Eq:Turbo_Eq38_7}
\gamma_{2,\, k,\, i,\, m,\, n,\, l} =
                            \exp
                            \left[-
                            \frac{
                            \left(
                            \tilde R_{k,\, j,\, l}-
                            \hat H_{k,\, j,\, l}
                             S_{m,\, n}
                            \right)^2}{2L_d\hat\sigma_w^2}
                            \right]
\end{eqnarray}
%*******************************************************************************
where
%*******************************************************************************
\begin{eqnarray}
\label{Eq:Turbo_Eq38_8}
j = L_{d2}/2 + i.
\end{eqnarray}
%*******************************************************************************
For a rate-1 turbo code, alternate gammas have to be set to unity, as
explained in the last paragraph of Section~\ref{SSec:Turbo_Dec}.
%*******************************************************************************
\subsection{The Channel Capacity}
\label{SSec:Channel_Cap}
%*******************************************************************************
The communication system model under consideration is given by
(\ref{Eq:Pap8_Eq38_3_1}). The channel capacity is given by \cite{Salehi_Dig}:
%*******************************************************************************
\begin{eqnarray}
\label{Eq:Channel_Cap_Eq1}
C = \frac{1}{2}
    \log_2(1+\mbox{SNR})
    \qquad \mbox{bits/transmission}
\end{eqnarray}
%*******************************************************************************
per dimension (real-valued signals occupy a single dimension, complex-valued
signals occupy two dimensions). The ``SNR'' in (\ref{Eq:Channel_Cap_Eq1})
denotes the minimum average signal-to-noise ratio per dimension, for
error-free transmission. Observe that:
%*******************************************************************************
\begin{enumerate}
 \item The sphere packing derivation of the channel capacity formula
       \cite{Salehi_Dig}, does not require noise to be Gaussian. The only
       requirements are that the noise samples have to be independent,
       the signal and noise have to be independent, and both the signal and
       noise must have zero mean.
 \item The channel capacity depends only on the SNR.
 \item The average SNR per dimension in (\ref{Eq:Channel_Cap_Eq1}) is
       different from the average
       SNR per bit (or $E_b/N_0$), which is widely used in the literature. In
       fact, it can be shown that \cite{Vasu_Book10,Salehi_Dig}:
%*******************************************************************************
\begin{eqnarray}
\label{Eq:Channel_Cap_Eq1_1}
\mbox{SNR} = 2C \times
\mbox{SNR per bit.}
\end{eqnarray}
%*******************************************************************************
 \item It is customary to define the average SNR per bit ($E_b/N_0$) over two
       dimensions (complex signals). When the signal and noise statistics over
       both dimensions are identical, the average SNR per bit over two
       dimensions is identical to the average SNR per bit over one dimension.
       Therefore (\ref{Eq:Channel_Cap_Eq1_1}) is valid, even though the
       SNR is defined over one dimension and the SNR per bit is defined
       over two dimensions.
 \item The notation $E_b/N_0$ is usually used for continuous-time, passband
       analog signals \cite{Haykin01,Proakis95,Salehi_Dig}, whereas SNR per
       bit is used for discrete-time signals \cite{Vasu_Book10}. However, both
       definitions are equivalent. Note that passband signals are capable of
       carrying information over two dimensions, using sine and cosine
       carriers, inspite of the fact that passband signals are real-valued.
 \item Each dimension corresponds to a separate and independent path between
       the transmitter and receiver.
 \item The channel capacity is additive with respect to the number of
       dimensions. Thus, the total capacity over $2N$ real dimensions is equal
       to the sum of the capacity over each real dimension.
 \item Each $S_{k,\, 3,\, i}$ in (\ref{Eq:Pap8_Eq38_3_1}) corresponds to
       one transmission (over two dimensions, since $S_{k,\, 3,\, i}$ is
       complex-valued).
 \item Transmission of $L_{d2}$ data bits in Figure~\ref{Fig:Enh_Frame} (for a
       rate-1 turbo code),
       results in $NL_{d2}$ complex samples ($2NL_{d2}$
       real-valued samples) of
       $\tilde R_{k,\, i,\, l}$ in (\ref{Eq:Pap8_Eq38_3_1}), for $N^{th}$-order
       receive diversity. Therefore, the channel capacity is
%*******************************************************************************
\begin{eqnarray}
\label{Eq:Channel_Cap_Eq2}
C & = & \frac{L_{d2}}{2NL_{d2}}                 \nonumber  \\
  & = & \frac{1}{2N}
        \qquad \mbox{bits/transmission}
\end{eqnarray}
%*******************************************************************************
       per dimension. In other words, (\ref{Eq:Channel_Cap_Eq2}) implies that
       in each transmission, one data bit is transmitted over $2N$ dimensions.
       Similarly, for a rate-$1/2$ turbo code with
       $N^{th}$-order receive diversity, transmission of
       $L_{d2}/2$ data bits results in $NL_{d2}$ complex samples of
       $\tilde R_{k,\, i,\, l}$ in (\ref{Eq:Pap8_Eq38_3_1}), and the channel
       capacity becomes:
%*******************************************************************************
\begin{eqnarray}
\label{Eq:Channel_Cap_Eq2_1}
C & = & \frac{L_{d2}}{4NL_{d2}}                 \nonumber  \\
  & = & \frac{1}{4N}
        \qquad \mbox{bits/transmission}
\end{eqnarray}
%*******************************************************************************
       per dimension. Substituting
       (\ref{Eq:Channel_Cap_Eq2}) and (\ref{Eq:Channel_Cap_Eq2_1}) in
       (\ref{Eq:Channel_Cap_Eq1}), and using (\ref{Eq:Channel_Cap_Eq1_1})
       we get the
       minimum (threshold) average SNR per bit required for error-free
       transmission, for a given channel capacity.
%*******************************************************************************
\begin{table}[tbh]
\centering
\caption{The minimum SNR per bit for different code rates and
         receiver diversity.}
\input{chan_cap.pstex_t}
\label{Tbl:Chan_Cap}
\end{table}
%*******************************************************************************
       The minimum SNR per bit for various code rates and receiver
       diversity is presented in Table~\ref{Tbl:Chan_Cap}. Note that
       \cite{Salehi_Dig} the minimum $E_b/N_0$ for
       error-free transmission is $-1.6$ dB only when $C\rightarrow 0$.
 \item In the case of fading channels, it may not be possible to achieve
       the minimum possible SNR per bit. This is because, the SNR per bit of a
       given frame may be less than the threshold average SNR per bit. Such
       frames are said to be in outage. The frame SNR per bit can be defined
       as (for the $k^{th}$ frame and the $l^{th}$ diversity arm):
%*******************************************************************************
\begin{eqnarray}
\label{Eq:Channel_Cap_Eq3}
\mbox{SNR}_{k,\, l,\,\mathrm{bit}} =
                      \frac{1}{2C}
                      \frac{<|\tilde H_{k,\, i,\, l} S_{k,\, 3,\, i}|^2>}
                           {<|\tilde W_{k,\, i,\, l}|^2>}
\end{eqnarray}
%*******************************************************************************
       where $<\cdot>$ denotes time average over the $L_{d2}$ data symbols.
       Note that the frame SNR is different from the average SNR per bit,
       which is defined as:
%*******************************************************************************
\begin{eqnarray}
\label{Eq:Channel_Cap_Eq4}
\mbox{SNR per bit} =
             \frac{1}{2C}
             \frac{E\left[\left|\tilde H_{k,\, i,\, l}
                                       S_{k,\, 3,\, i}
                    \right|^2\right]}
                  {E\left[\left|\tilde W_{k,\, i,\, l}\right|^2\right]}.
\end{eqnarray}
%*******************************************************************************
       The $k^{th}$ OFDM frame is said to be in outage when:
%*******************************************************************************
\begin{eqnarray}
\label{Eq:Channel_Cap_Eq5}
\mbox{SNR}_{k,\, l,\,\mathrm{bit}} <
\mbox{minimum average SNR per bit}
\end{eqnarray}
%*******************************************************************************
       for all $l$. The outage probability is given by:
%*******************************************************************************
\begin{eqnarray}
\label{Eq:Channel_Cap_Eq6}
P_{\mathrm{out}} = \frac{\mbox{number of frames in outage}}
                        {\mbox{total number of frames transmitted}}.
\end{eqnarray}
%*******************************************************************************
\end{enumerate}
%*******************************************************************************

%*******************************************************************************
\section{Simulation Results}
\label{Sec:Sim_Results}
\begin{figure}[tbh]
\centering
\input{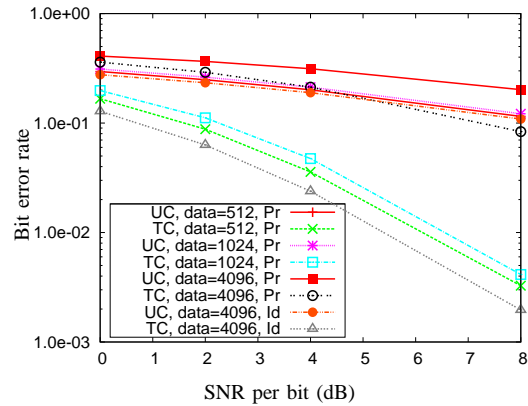}
\caption{Simulation results without data interleaving, frame structure in
         Figure~\ref{Fig:Frame}(a), rate-$1/2$ turbo code. \copyright\, 2013
         IEEE.  Reprinted, with permission, from \cite{Vasu13}.}
\label{Fig:DMT11_Ber}
\end{figure}

%*******************************************************************************
In this section, we present the simulation results for turbo-coded OFDM.
In the simulations, the channel length $L_h$ is equal to 10, hence $L_{hr}=19$.
The fade variance $\sigma^2_f=0.5$. The simulation results are presented in
Figure~\ref{Fig:DMT11_Ber}, for the frame structure in
Figure~\ref{Fig:Frame}(a) with $L_p=512$ and different values of $L_d$.
The term ``UC'' denotes uncoded, ``TC'' denotes turbo coded, ``data'' denotes
$L_{d1}$, ``Pr'' denotes practical receiver (with acquired synchronization and
channel estimates) and ``Id'' denotes ideal receiver (ideal synchronization
and channel estimates).
%*******************************************************************************
%*******************************************************************************

We find that for $L_{d1}=512$, the practical receiver has a performance that is
less than 1 dB inferior to the ideal receiver. However, the throughput of this
system is just 32.95\%, since the data length is equal to the preamble length.
Next, for $L_{d1}=1024$, the practical receiver is about 1 dB inferior to the
ideal receiver and the throughput
has improved to 39.72\%. When $L_{d1}=4096$, the performance of the practical
receiver is no better than
uncoded transmission. This is due to the fact that the residual RMS frequency
offset estimation error (fine) in Figure~\ref{Fig:DMT17_FOFF} is about
$2\times 10^{-4}$ radian, which is a significant fraction of the
subcarrier spacing ($2\pi/L_d=0.000767$ radian). Note that the frequency
offset estimation error depends only on $L_p$ and the performance of the
ideal receiver is independent of the data length $L_{d1}$.
%*******************************************************************************
\begin{figure}[tbh]
\centering
\input{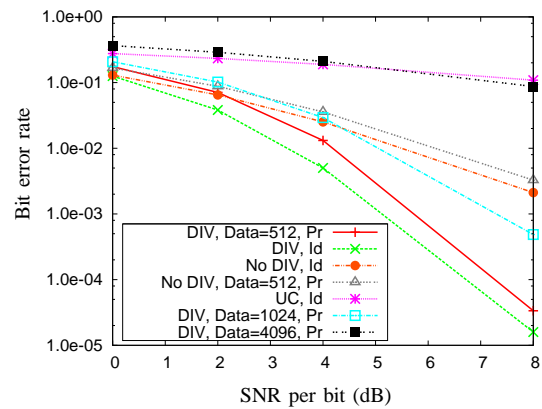}
\caption{Simulation results with data interleaving, frame structure in
         Figure~\ref{Fig:Frame}(a), rate-$1/2$ turbo code. \copyright\, 2013
         IEEE. Reprinted, with permission, from \cite{Vasu13}}
\label{Fig:DMT13_Ber}
\end{figure}
%*******************************************************************************

In Figure~\ref{Fig:DMT13_Ber}, we present the simulation results with
data interleaving, as discussed in Section~\ref{SSec:Data_Iv}. Again, the
performance of the ideal receiver is independent of $L_{d1}$.
We see that the practical receiver exhibits
more than two orders of magnitude improvement in the BER (compared
to the case where there is no data interleaving), at an SNR of 8 dB and
$L_{d1}=512$. When $L_{d1}$ is increased, the performance of the practical
receiver deteriorates.
%*******************************************************************************
\begin{figure}[tbh]
\centering
\input{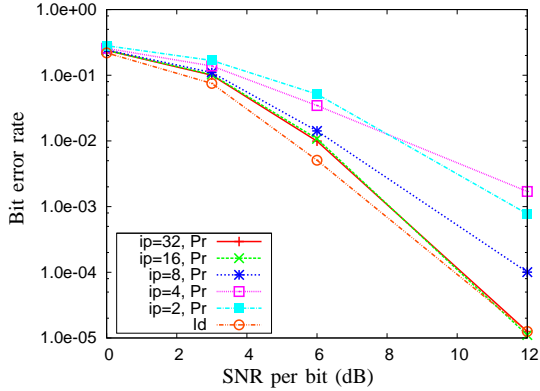}
\caption{Simulation results with data interleaving, enhanced frame structure
         in Figure~\ref{Fig:Enh_Frame}(a) and rate-1 turbo code.}
\label{Fig:DMT17_Ber}
\end{figure}
%*******************************************************************************

In Figure~\ref{Fig:DMT17_Ber}, we present simulation results for the
rate-1 turbo code, with enhanced frame structure, $1^{st}$-order
receiver diversity and interpolation factors (ip) equal to 2, 4, 8, 16 and 32.
We find that the performance of the practical receiver is as good as the
ideal receiver. However, there is a 4 dB degradation in performance of the
ideal receiver for the rate-1 turbo code,
with respect to the ideal receiver for the rate-$1/2$
turbo code in Figure~\ref{Fig:DMT13_Ber}, at a BER of $10^{-5}$. This
degradation in performance
can be compensated by using receiver diversity, which is presented next.
%*******************************************************************************
\begin{figure}[tbh]
\centering
\input{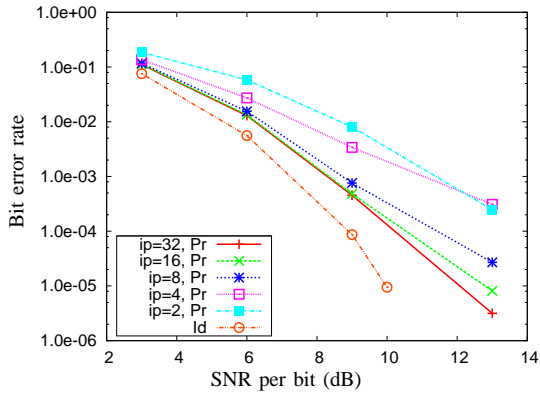}
\caption{Simulation results with data interleaving, enhanced frame structure
         in Figure~\ref{Fig:Enh_Frame}(a) and rate-1 turbo code with 2nd order
         receive diversity. Identical channel on both diversity arms.}
\label{Fig:DMT18_Ber}
\end{figure}
%*******************************************************************************

In Figure~\ref{Fig:DMT18_Ber}, we present simulation results for the
rate-1 turbo code, with enhanced frame structure and $2^{nd}$-order
receiver diversity. The channel in both diversity arms is assumed to be
identical. However, noise in both the diversity arms is assumed to be
independent. Comparing Figure~\ref{Fig:DMT17_Ber} and
Figure~\ref{Fig:DMT18_Ber}, we find that the ideal receiver with 2nd-order
diversity is just 2 dB better than the one with 1st-order diversity, at a
BER of $10^{-5}$. Moreover,
the practical receivers, with ip=32 have nearly identical performance. This
is to be expected, since it is well known that diversity advantage is obtained
only when the channels are independent.
%*******************************************************************************
\begin{figure}[tbh]
\centering
\input{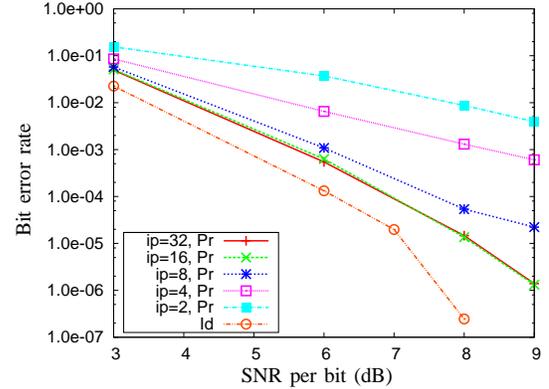}
\caption{Simulation results with data interleaving, enhanced frame structure
         in Figure~\ref{Fig:Enh_Frame}(a) and rate-1 turbo code with 2nd order
         receive diversity. Independent channel on both diversity arms.}
\label{Fig:DMT21_Ber}
\end{figure}
%*******************************************************************************

In Figure~\ref{Fig:DMT21_Ber}, we present simulation results for the
rate-1 turbo code, with enhanced frame structure and $2^{nd}$-order
receiver diversity. The channel and noise in both diversity arms are assumed
to be independent. Comparing Figure~\ref{Fig:DMT17_Ber} and
Figure~\ref{Fig:DMT21_Ber}, we find that the ideal receiver with 2nd order
diversity exhibits about 5 dB improvement over the one with 1st order
diversity, at a BER of $10^{-5}$. Moreover, the practical receiver with
ip=16, 32 is just 1 dB inferior to the ideal receiver, at a BER of $10^{-5}$.
%*******************************************************************************
\begin{figure}[tbh]
\centering
\input{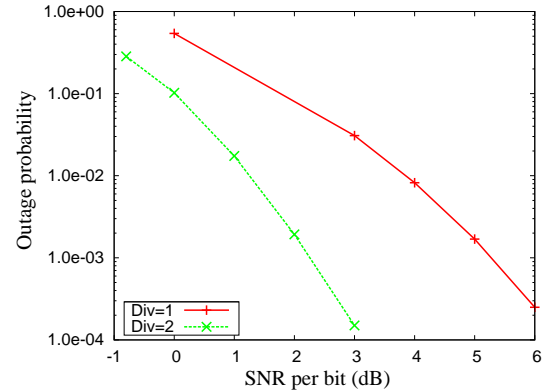}
\caption{Simulation results for outage probability with data interleaving,
         enhanced frame structure in Figure~\ref{Fig:Enh_Frame}(a) and rate-1
         turbo code with 1st and 2nd order receive diversity. Independent
         channel on both diversity arms.}
\label{Fig:DMT_Outage}
\end{figure}
%*******************************************************************************

Finally, in Figure~\ref{Fig:DMT_Outage} we present the outage probability
for the rate-1 turbo code with 1st and 2nd order receive diversity. The outage
probability for 1st order receive diversity, at 6 dB SNR per bit is
$3\times 10^{-4}$.
In other words, 3 out of $10^4$ frames are in outage (no error correcting code
can correct errors in such frames). Therefore, in the worst case, the
number of bit errors for the frames in outage would be
$0.5\times 3\times 3832$ (assuming probability of error is 0.5). Let us also
assume that for the remaining ($10000-3=9997$) frames, all errors are
corrected, using a sufficiently powerful error correcting code. Therefore,
in the best case situation, the overall BER at 6 dB SNR per bit, with 1st order
diversity would be $0.5\times 3\times 3832/(10000*3832)=1.5\times 10^{-4}$.
However, from Figure~\ref{Fig:DMT17_Ber}, even the ideal coherent receiver
exhibits a BER as high as $10^{-2}$ at 6 dB SNR per bit. Therefore, there is
large
scope for improvement, using perhaps a more powerful error correcting code.

Similarly we observe from Figure~\ref{Fig:DMT_Outage} that, with 2nd order
receive diversity, the outage probability is $10^{-4}$ at 3 dB SNR per bit.
This implies that 1 out of $10^{4}$ frames is in outage. Using similar
arguments, the best case overall BER at 3 dB SNR per bit would be
$0.5\times 3832/(10000*3832)=0.5\times 10^{-4}$. From
Figure~\ref{Fig:DMT21_Ber}, the ideal coherent receiver gives a BER of
$2\times 10^{-2}$, at 3 dB SNR per bit, once again suggesting that there
is large scope for improvement, using a better code.
%*******************************************************************************
\section{Conclusions and Future Work}
\label{Sec:Conclude}
%*******************************************************************************
This paper deals with linear complexity coherent detectors for turbo-coded OFDM
signals transmitted over frequency selective Rayleigh fading channels.
Simulation results show that it is possible to achieve a BER of $10^{-5}$
at an SNR per bit of 8 dB and throughput equal to 82.84\%, using a single
transmit and two receive antennas.

With the rapid advances in VLSI technology,
it is expected that coherent transceivers would drive the future wireless
telecommunication systems.

It may be possible to further improve the performance, using a better code.
%*******************************************************************************
%\section*{Acknowledgment}
%\label{Sec:Ack}
%*******************************************************************************
%This work is supported by the India-UK Advanced Technology Center (IU-ATC)
%of Excellence in Next Generation Networks, Systems and Services under grant
%SR/RCUK-DST/Next Gen(F)/2008, sponsored by DST-EPSRC.
%
% This work was presented as an Invited Talk
%at the International Federation of Nonlinear Analysts (IFNA)
%World Congress, Athens, Greece, 25th June--1st July, 2012.

\appendix
%*******************************************************************************
\subsection{An Approximate and Simple Cram{\' e}r-Rao Bound on the Variance of
            the Frequency Offset Estimation Error}
%*******************************************************************************
Consider the signal model in (\ref{Eq:Pap8_Eq5}), which is repeated here for
convenience (for notational simplicity, we drop the subscript $k$, assume
$\theta_k=0$ and $N-1=L_p-L_h+1$):
%*******************************************************************************
\begin{eqnarray}
\label{Eq:Ap_Eq1}
\tilde r_n = \tilde y_n
             \mathrm{e}^{\,\mathrm{j}\omega n} +
             \tilde w_n \quad \mbox{for $0\le n \le N-1$}.
\end{eqnarray}
%*******************************************************************************
We assume that the channel is known, and hence $\tilde y_n$ is known at the
receiver. Moreover, we consider only the steady-state preamble part of the
received signal (note that time is suitably re-indexed, such that the first
steady-state sample is considered as time zero, whereas, actually the first
steady-state sample occurs at time $L_h-1$). Define
%*******************************************************************************
\begin{eqnarray}
\label{Eq:Ap_Eq2}
\tilde\mathbf{y} & = &
\left[
\begin{array}{ccc}
\tilde y_0 & \ldots & \tilde y_{N-1}
\end{array}
\right]                                    \nonumber  \\
\tilde\mathbf{r} & = &
\left[
\begin{array}{ccc}
\tilde r_0 & \ldots & \tilde r_{N-1}
\end{array}
\right].
\end{eqnarray}
%*******************************************************************************

The coherent maximum likelihood (ML)
estimate of the frequency offset is obtained as follows: choose that value of
$\hat\omega$ which maximizes the joint conditional pdf
%*******************************************************************************
\begin{eqnarray}
\label{Eq:Ap_Eq3}
\max_{\hat\omega \in [-\omega_{\mathrm{max}},\, \omega_{\mathrm{max}}]}
 p
\left(
\tilde\mathbf{r}|\tilde\mathbf{y},\,\hat\omega
\right)
\end{eqnarray}
%*******************************************************************************
where $\omega_{\mathrm{max}}$ denotes the maximum possible frequency offset
in radians. Substituting for the joint conditional pdf in (\ref{Eq:Ap_Eq3}),
we obtain
%*******************************************************************************
\begin{eqnarray}
\label{Eq:Ap_Eq4}
\max_{\hat\omega}
\frac{1}{(2\pi\sigma^2_w)^{N}}
\exp
\left(
\frac{
-
\sum_{n=0}^{N-1}
\left|
\tilde r_n-\tilde y_n
\,
\mathrm{e}^{\,\mathrm{j}\,\hat\omega n}
\right|^2
}{2\sigma^2_w}
\right)
\end{eqnarray}
%*******************************************************************************
which simplifies to
%*******************************************************************************
\begin{eqnarray}
\label{Eq:Ap_Eq4_1}
\max_{\hat\omega}
\Re
\left
\{
\sum_{n=0}^{N-1}
\tilde r_n\tilde y_n^*
\,
\mathrm{e}^{-\mathrm{j}\,\hat\omega n}
\right
\}.
\end{eqnarray}
%*******************************************************************************
Observe that (\ref{Eq:Pap8_Eq31}) is the non-coherent ML frequency offset
(and timing) estimator, whereas (\ref{Eq:Ap_Eq4_1}) is the coherent ML
frequency offset estimator assuming timing is known.

Since ML estimators are unbiased, the variance of the frequency offset
estimate is lower bounded by the Cram{\' e}r-Rao bound (CRB):
%*******************************************************************************
\begin{eqnarray}
\label{Eq:Ap_Eq5}
E
\left[
\left(
\hat\omega - \omega
\right)^2
\right] \ge 1\bigg/
            E
            \left[
            \left(
            \frac{\partial}{\partial \omega}
            \ln p
            \left(
            \tilde\mathbf{r}|\tilde\mathbf{y},\,\omega
            \right)
            \right)^2
            \right]
\end{eqnarray}
%*******************************************************************************
since $\tilde\mathbf{y}$ is assumed to be known. It can be shown that
%*******************************************************************************
\begin{eqnarray}
\label{Eq:Ap_Eq6}
\frac{\partial}{\partial \omega}
\ln p
\left(
\tilde\mathbf{r}|\tilde\mathbf{y},\,\omega
\right)  & = & \frac{\mathrm{j}}{2\sigma^2_w}
               \sum_{n=0}^{N-1}
               \left[
                n
               \tilde y_n
               \mathrm{e}^{\,\mathrm{j}\,\omega n} \tilde w_n^*
               \right.                             \nonumber  \\
         &   & \left.
               \mbox{ } - n
               \tilde y_n^*
               \mathrm{e}^{-\mathrm{j}\,\omega n} \tilde w_n
               \right].
\end{eqnarray}
%*******************************************************************************
Substituting (\ref{Eq:Ap_Eq6}) in (\ref{Eq:Ap_Eq5}) and assuming independent
noise (the real and imaginary parts of noise are also assumed independent),
we obtain:
%*******************************************************************************
\begin{eqnarray}
\label{Eq:Ap_Eq7}
 E
\left[
\left(
\frac{\partial}{\partial \omega}
\ln p
\left(
\tilde\mathbf{r}|\tilde\mathbf{y},\,\omega
\right)
\right)^2
\right] = \frac{1}{\sigma^2_w}
          \sum_{n=0}^{N-1} n^2
          \left|
          \tilde y_n
          \right|^2
\end{eqnarray}
%*******************************************************************************
and hence
%*******************************************************************************
\begin{eqnarray}
\label{Eq:Ap_Eq8}
E
\left[
\left(
\hat\omega - \omega
\right)^2
\right] \ge \left[
            \frac{1}{\sigma^2_w}
            \sum_{n=0}^{N-1} n^2
            \left|
            \tilde y_n
            \right|^2
            \right]^{-1}
\end{eqnarray}
%*******************************************************************************
when $\tilde y_n$ is known. When $\tilde y_n$ is a random variable, which is
true in our case, then the right hand side of (\ref{Eq:Ap_Eq8}) needs to
be further averaged over $\tilde\mathbf{y}$ \cite{Mengali_2000,Zeng_2010}.
In other words, we need to compute
%*******************************************************************************
\begin{eqnarray}
\label{Eq:Ap_Eq9}
\lefteqn{
 E
\left[
\left(
\frac{1}{\sigma^2_w}
\sum_{n=0}^{N-1} n^2
\left|
\tilde y_n
\right|^2
\right)^{-1}
\right]}       \nonumber  \\
& = &
\int_{\tilde\mathbf{y}}
\left[
\frac{1}{\sigma^2_w}
\sum_{n=0}^{N-1} n^2
\left|
\tilde y_n
\right|^2
\right]^{-1}
p(\tilde\mathbf{y})\, d\tilde\mathbf{y}
\end{eqnarray}
%*******************************************************************************
which is complicated. The purpose of this Appendix is to provide an alternate
and a much simpler solution to (\ref{Eq:Ap_Eq9}), by assuming that
$\tilde y_n$ is ergodic.

We claim that, for large values of $N$ (in our case $N=504$)
%*******************************************************************************
\begin{eqnarray}
\label{Eq:Ap_Eq10}
\sum_{n=0}^{N-1} n^2
\left|
\tilde y_n
\right|^2
& \approx &
\sum_{n=0}^{N-1} n^2
 E
\left[
\left|
\tilde y_n
\right|^2
\right]                          \nonumber  \\
& = & \mbox{a constant}.
\end{eqnarray}
%*******************************************************************************
Now
%*******************************************************************************
\begin{eqnarray}
\label{Eq:Ap_Eq11}
\tilde y_n = \sum_{i=0}^{L_h-1}
             \tilde h_i \tilde s_{n-i}.
\end{eqnarray}
%*******************************************************************************
Therefore
%*******************************************************************************
\begin{eqnarray}
\label{Eq:Ap_Eq12}
 E
\left[
\left|
\tilde y_n
\right|^2
\right] & = &  E
              \left[
              \sum_{i=0}^{L_h-1}
              \tilde h_i \tilde s_{n-i}
              \sum_{j=0}^{L_h-1}
              \tilde h_j^* \tilde s_{n-j}^*
              \right]                            \nonumber  \\
        & = & \sum_{i=0}^{L_h-1}
              \sum_{j=0}^{L_h-1}
               E
              \left[
              \tilde h_i
              \tilde h_j^*
              \right]
               E
              \left[
              \tilde s_{n-i}
              \tilde s_{n-j}^*
              \right]
\end{eqnarray}
%*******************************************************************************
where we have assumed
%*******************************************************************************
\begin{enumerate}
 \item $\tilde h_n$ and $\tilde s_n$ to be independent
 \item $\tilde s_n$ (the preamble) varies randomly from frame to frame and is
       not a constant.
\end{enumerate}
%*******************************************************************************
Hence (\ref{Eq:Ap_Eq12}) can be rewritten as:
%*******************************************************************************
\begin{eqnarray}
\label{Eq:Ap_Eq13}
 E
\left[
\left|
\tilde y_n
\right|^2
\right] & = & \sum_{i=0}^{L_h-1}
              \sum_{j=0}^{L_h-1}
               2
              \sigma^2_f \delta_K(i-j)
              \sigma^2_s \delta_K(j-i)                 \nonumber  \\
        & = &  2
              \sigma^2_f
              \sigma^2_s L_h.
\end{eqnarray}
%*******************************************************************************
where $\sigma^2_f$ is defined in (\ref{Eq:Pap8_Eq3}), $\sigma^2_s$ is defined
in (\ref{Eq:Pap8_Eq4_2}) and $\delta_K(\cdot)$ is the Kronecker delta
function. With these developments (\ref{Eq:Ap_Eq9}) becomes
%*******************************************************************************
\begin{eqnarray}
\label{Eq:Ap_Eq14}
 E
\left[
\left(
\frac{1}{\sigma^2_w}
\sum_{n=0}^{N-1} n^2
\left|
\tilde y_n
\right|^2
\right)^{-1}
\right]
\approx
\left[
\frac{2\sigma^2_f\sigma^2_s L_h}{\sigma^2_w}
\sum_{n=0}^{N-1} n^2
\right]^{-1}.
\end{eqnarray}
%*******************************************************************************
Therefore, the CRB on the variance of the frequency offset estimate is
(assuming $N-1=M$)
%*******************************************************************************
\begin{eqnarray}
\label{Eq:Ap_Eq15}
E
\left[
\left(
\hat\omega - \omega
\right)^2
\right] \ge \left[
            \frac{2\sigma^2_f\sigma^2_s L_h}{\sigma^2_w}
            \left(
            \frac{M^3}{3} +
            \frac{M^2}{2} +
            \frac{M}{6}
            \right)
            \right]^{-1}
\end{eqnarray}
%*******************************************************************************

%*******************************************************************************
\bibliographystyle{IEEEtran}
\bibliography{/home/vasu/vasu/bib/mybib,/home/vasu/vasu/bib/mybib1,/home/vasu/vasu/bib/mybib2,/home/vasu/vasu/bib/mybib3,/home/vasu/vasu/bib/mybib4.bib}

% Generated by IEEEtran.bst, version: 1.12 (2007/01/11)
\begin{thebibliography}{10}
\providecommand{\url}[1]{#1}
\csname url@samestyle\endcsname
\providecommand{\newblock}{\relax}
\providecommand{\bibinfo}[2]{#2}
\providecommand{\BIBentrySTDinterwordspacing}{\spaceskip=0pt\relax}
\providecommand{\BIBentryALTinterwordstretchfactor}{4}
\providecommand{\BIBentryALTinterwordspacing}{\spaceskip=\fontdimen2\font plus
\BIBentryALTinterwordstretchfactor\fontdimen3\font minus
  \fontdimen4\font\relax}
\providecommand{\BIBforeignlanguage}[2]{{%
\expandafter\ifx\csname l@#1\endcsname\relax
\typeout{** WARNING: IEEEtran.bst: No hyphenation pattern has been}%
\typeout{** loaded for the language `#1'. Using the pattern for}%
\typeout{** the default language instead.}%
\else
\language=\csname l@#1\endcsname
\fi
#2}}
\providecommand{\BIBdecl}{\relax}
\BIBdecl

\bibitem{Hanzo_2011}
L.~Hanzo, M.~El-Hajjar, and O.~Alamri, ``{Near-Capacity Wireless Transceivers
  and Cooperative Communications in the MIMO Era: Evolution of Standards,
  Waveform Design, and Future Perspectives},'' \emph{Proc. IEEE}, vol.~99,
  no.~8, pp. 1343--1385, Aug. 2011.

\bibitem{Zhang_2013}
R.~Zhang \emph{et~al.}, ``{Advances in Base- and Mobile-Station Aided
  Cooperative Wireless Communications},'' \emph{IEEE Veh. Tech. Mag.}, vol.~8,
  no.~1, pp. 57--69, March 2013.

\bibitem{Hanzo_2012}
L.~Hanzo, H.~Haas, S.~Imre, D.~O. Brien, M.~Rupp, and L.~Gyongyosi, ``{Wireless
  Myths, Realities, and Futures: From 3G/4G to Optical and Quantum Wireless},''
  \emph{Proc. IEEE}, vol. 100, no. Special Centennial issue, pp. 1853--1888,
  May 2012.

\bibitem{Vasu13}
K.~Vasudevan, ``{Coherent Detection of Turbo Coded OFDM Signals Transmitted
  through Frequency Selective Rayleigh Fading Channels},'' in \emph{Proc. IEEE
  ISPCC, Shimla, India}, Sept. 2013.

\bibitem{Umesh13}
U.~C. Samal and K.~Vasudevan, ``{Bandwidth Efficient Turbo Coded OFDM
  Systems},'' in \emph{Proc. IEEE ITST, Tampere, Finland}, Nov. 2013, pp.
  490--495.

\bibitem{Bingham90}
J.~A.~C. Bingham, ``{Multicarrier Modulation for Data Transmission: An Idea
  Whose Time Has Come},'' \emph{IEEE Commun. Mag.}, vol.~28, no.~5, pp. 5--14,
  May 1990.

\bibitem{Vasu_Book10}
K.~Vasudevan, \emph{{Digital Communications and Signal Processing, Second
  edition (CDROM included)}}.\hskip 1em plus 0.5em minus 0.4em\relax
  Universities Press (India), Hyderabad, www.universitiespress.com, 2010.

\bibitem{Hanzo_OFDM_Primer}
L.~Hanzo and T.~Keller, \emph{{OFDM and MC-CDMA: A Primer}}.\hskip 1em plus
  0.5em minus 0.4em\relax John Wiley, 2006.

\bibitem{Cox97}
T.~M. Schmidl and D.~C. Cox, ``{Robust Frequency and Timing Synchronization for
  OFDM},'' \emph{IEEE Trans. on Commun.}, vol.~45, no.~12, pp. 1613--1621, Dec.
  1997.

\bibitem{Beek95_2}
J.-J. van~de Beek, M.~Sandell, M.~Isaksson, and P.~O. B{\" o}rjesson,
  ``{Low-Complex Frame Synchronization in OFDM Systems},'' in \emph{Proc. of
  the $4^{th}$ IEEE International Conference on Universal Personal
  Communications}, Nov. 1995, pp. 982--986.

\bibitem{Landstrom02}
D.~Landstr{\" o}m, S.~K. Wilson, J.-J. van~de Beek, P.~{\" O}dling, and P.~O.
  B{\" o}rjesson, ``{Symbol Time Offset Estimation in Coherent OFDM Systems},''
  \emph{IEEE Trans. on Commun.}, vol.~50, no.~4, pp. 545--549, April 2002.

\bibitem{Cheon03}
B.~Park, H.~Cheon, C.~Kang, and D.~Hong, ``{A Novel Timing Estimation Method
  for OFDM Systems},'' \emph{IEEE Commun. Lett.}, vol.~7, no.~5, pp. 239--241,
  May 2003.

\bibitem{Ren05}
G.~Ren, Y.~Chang, H.~Zhang, and H.~Zhang, ``{Synchronization Method Based on a
  New Constant Envelope Preamble for OFDM Systems},'' \emph{IEEE Trans. on
  Broadcasting}, vol.~51, no.~1, pp. 139--143, Mar. 2005.

\bibitem{Kang08}
Y.~Kang, S.~Kim, D.~Ahn, and H.~Lee, ``{Timing Estimation for OFDM Systems by
  using a Correlation Sequence of Preamble},'' \emph{IEEE Trans. on Consumer
  Electronics}, vol.~54, no.~4, pp. 1600--1608, Nov. 2008.

\bibitem{Baum98}
K.~L. Baum, ``{A Synchronous Coherent OFDM Air Interface Concept for High Data
  Rate Cellular Systems},'' in \emph{Proc. of the IEEE VTS 48th Vehicular
  Technology Conf.}, May 1998, pp. 2222--2226.

\bibitem{Garcia01}
M.~Julia Fern{\' a}ndez-Getino~Garcia, O.~Edfors, and J.~M. P{\' a}ez-Borrallo,
  ``{Frequency Offset Correction for Coherent OFDM in Wireless Systems},''
  \emph{IEEE Trans. on Consumer Electronics}, vol.~47, no.~1, pp. 187--193,
  Feb. 2001.

\bibitem{Bradaric03}
I.~Bradaric and A.~P. Petropulu, ``{Blind Estimation of the Carrier Frequency
  Offset in OFDM Systems},'' in \emph{4th IEEE Workshop on Signal Processing
  Advances in Wireless Communications}, June 2003, pp. 590--594.

\bibitem{Kuo05}
C.~Kuo and J.-F. Chang, ``{Generalized Frequency Offset Estimation in OFDM
  Systems Using Periodic Training Symbol},'' in \emph{Proc. IEEE Intl. Conf. on
  Commun.}, May 2005, pp. 715--719.

\bibitem{Lin_Chen_05}
J.-S. Lin and C.-C. Chen, ``{Hybrid Maximum Likelihood Frequency Offset
  Estimation in Coherent OFDM Systems},'' \emph{IEE Proc.--Commun.}, vol. 152,
  no.~5, pp. 587--592, Oct. 2005.

\bibitem{Ahn07}
S.~Ahn, C.~Lee, S.~Kim, S.~Yoon, and S.~Y. Kim, ``{A Novel Scheme for Frequency
  Offset Estimation in OFDM Systems},'' in \emph{9th Intl. Conf. on Adv.
  Commun. Technol.}, Feb. 2007, pp. 1632--1635.

\bibitem{Henkel07}
M.~Henkel and W.~Schroer, ``{Pilot Based Synchronization Strategy for a
  Coherent OFDM Receiver},'' in \emph{IEEE Wireless Communications and
  Networking Conference (WCNC)}, March 2007, pp. 1984--1988.

\bibitem{Tufvesson99}
F.~Tufvesson, O.~Edfors, and M.~Faulkner, ``{Time and Frequency Synchronization
  for OFDM using PN-sequence Preambles},'' in \emph{Proc. of the IEEE VTS 50th
  Vehicular Technology Conf.}, Sept. 1999, pp. 2203--2207.

\bibitem{Minn03}
H.~Minn, V.~K. Bhargava, and K.~B. Letaief, ``{A Robust Timing and Frequency
  Synchronization for OFDM Systems},'' \emph{IEEE Trans. on Wireless Commun.},
  vol.~2, no.~4, pp. 822--839, July 2003.

\bibitem{Zhang05}
Z.~Zhang, K.~Long, M.~Zhao, and Y.~Liu, ``{Joint Frame Synchronization and
  Frequency Offset Estimation in OFDM Systems},'' \emph{IEEE Trans. on
  Broadcasting}, vol.~51, no.~3, pp. 389--394, Sept. 2005.

\bibitem{Ziabari2011}
H.~Abdzadeh-Ziabari and M.~G. Shayesteh, ``{Robust Timing and Frequency
  Synchronization for OFDM Systems},'' \emph{IEEE Trans. on Veh. Technol.},
  vol.~60, no.~8, pp. 3646--3656, Oct. 2011.

\bibitem{Tanda2013}
D.~Mattera and M.~Tanda, ``{Blind Symbol Timing and CFO Estimation for
  OFDM/OQAM Systems},'' \emph{IEEE Trans. on Wireless Commun.}, vol.~12, no.~1,
  pp. 268--277, Jan. 2013.

\bibitem{Salcedo2013}
J.~A. L{\' o}pez-Salcedo, E.~Guti{\' e}rrez, G.~Seco-Granados, and A.~L.
  Swindlehurst, ``{Unified Framework for the Synchronization of Flexible
  Multicarrier Communication Signals},'' \emph{IEEE Trans. on Sig. Proc.},
  vol.~61, no.~4, pp. 828--842, Feb. 2013.

\bibitem{Frenger99}
P.~K. Frenger and N.~A.~B. Svensson, ``{Decision-Directed Coherent Detection in
  Multicarrier Systems on Rayleigh Fading Channels},'' \emph{IEEE Trans. on
  Veh. Technol.}, vol.~48, no.~2, pp. 490--498, Mar. 1999.

\bibitem{Merli08}
F.~Z. Merli and G.~M. Vitetta, ``{A Factor Graph Approach to the Iterative
  Detection of OFDM Signals in the Presence of Carrier Frequency Offset and
  Phase Noise},'' \emph{IEEE Trans. on Wireless Commun.}, vol.~7, no.~3, pp.
  868--877, Mar. 2008.

\bibitem{Wang12}
H.~wei Wang, D.~W. Lin, and T.-H. Sang, ``{OFDM Signal Detection in Doubly
  Selective Channels with Blockwise Whitening of Residual Intercarrier
  Interference and Noise},'' \emph{IEEE J. on Select. Areas in Commun.},
  vol.~30, no.~4, pp. 684--694, May 2012.

\bibitem{Chen2013}
C.-Y. Chen, Y.-Y. Lan, and T.-D. Chiueh, ``{Turbo Receiver with ICI-Aware
  Dual-List Detection for Mobile MIMO-OFDM Systems},'' \emph{IEEE Trans. on
  Wireless Commun.}, vol.~12, no.~1, pp. 100--111, Jan. 2013.

\bibitem{Beek95}
J.-J. van~de Beek, O.~Edfors, M.~Sandell, S.~K. Wilson, and P.~O. B{\"
  o}rjesson, ``{On Channel Estimation in OFDM Systems},'' in \emph{Proc. of the
  IEEE VTS 45th Vehicular Technology Conf.}, July 1995, pp. 815--819.

\bibitem{Edfors98}
O.~Edfors, M.~Sandell, J.-J. van~de Beek, S.~K. Wilson, and P.~O. B{\"
  o}rjesson, ``{OFDM Channel Estimation by Singular Value Decomposition},''
  \emph{IEEE Trans. on Commun.}, vol.~46, no.~7, pp. 931--939, July 1998.

\bibitem{Puri02}
S.~Coleri, M.~Ergen, A.~Puri, and A.~Bahai, ``{Channel Estimation Techniques
  Based on Pilot Arrangement in OFDM Systems},'' \emph{IEEE Trans. on
  Broadcasting}, vol.~48, no.~3, pp. 223--229, Sept. 2002.

\bibitem{Ribeiro08}
C.~Ribeiro and A.~Gameiro, ``{An OFDM Symbol Design for Reduced Complexity MMSE
  Channel Estimation},'' \emph{Journal of Communications, Academy Publisher},
  vol.~3, no.~4, pp. 26--33, Sept. 2008.

\bibitem{Kinjo08}
S.~Kinjo, ``{An MMSE Channel Estimation Algorithm Based on the Conjugate
  Gradient Method for OFDM Systems},'' in \emph{The 23rd International
  Technical Conference on Circuits/Systems, Computers and Communications
  (ITC-CSCC)}, July 2008, pp. 969--972.

\bibitem{Jiang07}
M.~Jiang, J.~Akhtman, and L.~Hanzo, ``{Iterative Joint Channel Estimation and
  Multi-User Detection for Multiple-Antenna Aided OFDM Systems},'' \emph{IEEE
  Trans. on Wireless Commun.}, vol.~6, no.~8, pp. 2904--2914, Aug. 2007.

\bibitem{Fischer01}
R.~F.~H. Fischer, L.~H.-J. Lampe, and S.~H. M{\" u}ller-Weinfurtner, ``{Coded
  Modulation for Noncoherent Reception with Application to OFDM},'' \emph{IEEE
  Trans. on Veh. Technol.}, vol.~50, no.~4, pp. 910--919, July 2001.

\bibitem{Marey12}
M.~Marey, M.~Samir, and O.~A. Dobre, ``{EM-Based Joint Channel Estimation and
  IQ Imbalances for OFDM Systems},'' \emph{IEEE Trans. on Broadcasting},
  vol.~58, no.~1, pp. 106--113, Mar. 2012.

\bibitem{Kamalian12}
M.~Kamalian, A.~A. Tadaion, and M.~Derakhtian, ``{Invariant Detection of
  Orthogonal Frequency Division Multiplexing Signals with Unknown Parameters
  for Cognitive Radio Applications},'' \emph{IET Sig. Proc.}, vol.~6, no.~3,
  pp. 205--212, 2012.

\bibitem{Turunen12}
V.~Turunen, M.~Kosunen, M.~V{\" a }{\" a }r{\" a }kangas, and J.~Ryyn{\" a
  }nen, ``{Correlation-Based Detection of OFDM Signals in the Angular
  Domain},'' \emph{IEEE Trans. on Veh. Technol.}, vol.~61, no.~3, pp. 951--958,
  Mar. 2012.

\bibitem{Peng06}
P.~Tan and N.~C. Beaulieu, ``{A Comparison of DCT-Based OFDM and DFT-Based OFDM
  in Frequency Offset and Fading Channels},'' \emph{IEEE Trans. on Commun.},
  vol.~54, no.~11, pp. 2113--2125, Nov. 2006.

\bibitem{Yu12}
C.~Yu, C.-H. Sung, C.-H. Kuo, M.-H. Yen, and S.-J. Chen, ``{Design and
  Implementation of a Low-Power OFDM Receiver for Wireless Communications},''
  \emph{IEEE Trans. on Consumer Electronics}, vol.~58, no.~3, pp. 739--745,
  Aug. 2012.

\bibitem{Jeanclaude95}
H.~Sari, G.~Karam, and I.~Jeanclaude, ``{Transmission Techniques for Digital
  Terrestrial TV Broadcasting},'' \emph{IEEE Commun. Mag.}, vol.~33, no.~2, pp.
  100--109, Feb. 1995.

\bibitem{Reimers98}
U.~Reimers, ``{Digital Video Broadcasting},'' \emph{IEEE Commun. Mag.},
  vol.~36, no.~6, pp. 104--110, June 1998.

\bibitem{Takahashi09}
H.~Takahashi, ``{Coherent OFDM Transmission with High Spectral Efficiency},''
  in \emph{35th European Conference on Optical Communication}, Sept. 2009, pp.
  1--4.

\bibitem{Vasu08}
K.~Vasudevan, ``{Synchronization of Bursty Offset QPSK Signals in the Presence
  of Frequency Offset and Noise},'' in \emph{Proc. IEEE TENCON, Hyderabad,
  India}, Nov. 2008.

\bibitem{Vasu_SIVP10}
------, ``{Iterative Detection of Turbo Coded Offset QPSK in the Presence of
  Frequency and Clock Offsets and AWGN},'' \emph{Signal, Image and Video
  Processing, Springer}, vol.~6, no.~4, pp. 557--567, Nov. 2012.

\bibitem{Haykin_Adapt_96}
S.~Haykin, \emph{{Adaptive Filter Theory}}, 3rd~ed.\hskip 1em plus 0.5em minus
  0.4em\relax Prentice Hall, 1996.

\bibitem{Baxley10}
R.~J. Baxley, B.~T. Walkenhorst, and G.~Acosta-Marum, ``{Complex Gaussian Ratio
  Distribution with Applications for Error Rate Calculation in Fading Channels
  with Imperfect CSI},'' in \emph{Proc. IEEE Global Telecomm. Conf.}, Dec.
  2010, pp. 1--5.

\bibitem{Torres10}
J.~J. S{\' a}nchez-S{\' a}nchez, U.~Fern{\' a}ndez-Plazaola, and
  M.~Aguayo-Torres, ``{Sum of Ratios of Complex Gaussian RVs and its
  Application to a Simple OFDM Relay Network},'' in \emph{Proc. of the IEEE VTS
  71st Vehicular Technology Conf.}, May 2010, pp. 1--5.

\bibitem{Singer04}
R.~Koetter, A.~C. Singer, and M.~T{\" u}chler, ``{Turbo Equalization},''
  \emph{IEEE Sig. Proc. Mag.}, vol.~21, no.~1, pp. 67--80, Jan. 2004.

\bibitem{Darshan13}
D.~V. Adakane and K.~Vasudevan, ``{An Efficient Pilot Pattern Design for
  Channel Estimation in OFDM Systems},'' in \emph{Proc. IEEE ISPCC, Shimla,
  India}, Sept. 2013.

\bibitem{Salehi_Dig}
J.~G. Proakis and M.~Salehi, \emph{{Fundamentals of Communication
  Systems}}.\hskip 1em plus 0.5em minus 0.4em\relax Pearson Education Inc.,
  2005.

\bibitem{Haykin01}
S.~Haykin, \emph{{Communication Systems}}, 4th~ed.\hskip 1em plus 0.5em minus
  0.4em\relax Wiley Eastern, 2001.

\bibitem{Proakis95}
J.~G. Proakis, \emph{{Digital Communications}}, 3rd~ed.\hskip 1em plus 0.5em
  minus 0.4em\relax McGraw Hill, 1995.

\bibitem{Mengali_2000}
M.~Morelli and U.~Mengali, ``{Carrier-Frequency Estimation for Transmissions
  over Selective Channels},'' \emph{IEEE Trans. on Commun.}, vol.~48, no.~9,
  pp. 1580--1589, Sept. 2000.

\bibitem{Zeng_2010}
Y.~Li, H.~Minn, and J.~Zeng, ``{An Average Cramer-Rao Bound for Frequency
  Offset Estimation in Frequency-Selective Fading Channels},'' \emph{IEEE
  Trans. on Wireless Commun.}, vol.~9, no.~3, pp. 871--875, March 2010.

\end{thebibliography}
\end{document}